\documentclass[twocolumn, aps, prresearch, superscriptaddress, longbibliography]{revtex4-2}

\usepackage[utf8]{inputenc}
\usepackage[T1]{fontenc}
\usepackage[english]{babel}
\usepackage{amsmath,amssymb,amsfonts,amsthm,mathtools}
\usepackage{physics,bm,braket}
\usepackage{graphicx}
\usepackage{xcolor}
\usepackage{listings}
\usepackage{wrapfig}
\usepackage{placeins}
\usepackage{algorithm}
\usepackage{algpseudocode}
\usepackage[inline]{enumitem}
\usepackage{tikz}
\usepackage{algorithm}
\usepackage{algpseudocode}
\usepackage{amsmath}
\usepackage{svg}
\usepackage{natbib}
\usepackage{booktabs}
\usetikzlibrary{arrows.meta}
\usepackage{tensor-network}
\tnsetup{tensorsize=22pt, xscale=1.5, yscale=0.9, outer ysep=7pt}
\tnsetup{tensor fill=blue!12, tensor draw=blue!70!black, tensor text=blue!45!black, tensor opacity=0.9, shade top=orange!10, shade bottom=orange!70}
\usetikzlibrary{positioning}

\lstdefinelanguage{Julia}%
  {morekeywords={abstract,break,case,catch,const,continue,do,else,elseif,%
      end,export,false,for,function,immutable,import,importall,if,in,%
      macro,module,otherwise,quote,return,switch,true,try,type,typealias,%
      using,while},%
   sensitive=true,%
   alsoother={},%
   morecomment=[l]\#,%
   morecomment=[n]{\#=}{=\#},%
   morestring=[s]{"}{"},%
   morestring=[m]{'}{'},%
}[keywords,comments,strings]%

\DeclareUnicodeCharacter{2212}{\ensuremath{-}}

\definecolor{deeppurple}{rgb}{0.7, 0, 0.8}

\definecolor{tensorblue}{RGB}{171,206,231}
\definecolor{tensorbluee}{RGB}{128,184,219}
\definecolor{tensorred}{RGB}{245,170,170}
\definecolor{tensorredd}{RGB}{240,130,136}
\definecolor{tensorpurp}{rgb}{1,0.5,1}
\definecolor{tensorgreen}{RGB}{200,205,208}
\definecolor{uhhblue}{RGB}{2,113,187}
\definecolor{uhhgrey}{RGB}{59,81,91}
\definecolor{uhhdarkgrey}{RGB}{17,41,47}
\definecolor{uhhred}{RGB}{226,0,26}
\definecolor{uhhdarkred}{RGB}{174,3,29}
\definecolor{uhhdarkblue}{RGB}{0,78,129}
\definecolor{luhhblue}{RGB}{209,226,241}
\definecolor{luhhgrey}{RGB}{222,225,227}
\definecolor{luhhred}{RGB}{250,208,205}

\tikzset{ten/.style={fill=tensorblue}}
\tikzset{tenred/.style={fill=tensorred}}
\tikzset{tengreen/.style={fill=tensorgreen}}
\tikzset{tenpurp/.style={fill=tensorpurp}}
\tikzset{tengrey/.style={fill=black!20}}
\tikzset{tenorange/.style={fill=orange!30}}
\tikzset{u/.style={fill=blue!20,draw=black}}
\tikzset{w/.style={fill=green!50!black!80,draw=black}}

\usetikzlibrary{decorations.markings}
\usetikzlibrary{decorations.pathreplacing}
\usetikzlibrary{arrows.meta}
\usetikzlibrary{calc}
\tikzset{
  ->-/.style={postaction={decorate}, decoration={markings, mark=at position #1 with {\arrow{>}}}},
  <-/.style={postaction={decorate}, decoration={markings, mark=at position #1 with {\arrow{<}}}},
}

\allowdisplaybreaks
\advance\parskip 0.4pt

\newif\ifproduction
\productionfalse %

\usepackage{hyperref}
\hypersetup{
	colorlinks=true,
	linkcolor=blue,
	filecolor=blue,
	citecolor = red,      
	urlcolor=cyan,
}%

\newcommand{\Ra}{\mathrm{Ra}}   %
\newcommand{\Pran}{\mathrm{Pr}} %
\newcommand{\Nu}{\mathrm{Nu}}   %
\newcommand{\Nuens}{\overline{\Nu}}

\newcommand{\ff}{\mathrm{FF}}
\newcommand{\dns}{\text{DNS}}
\newcommand{\mps}{\text{MPS}}
\newcommand{\DT}{\mathcal{T}}
\usepackage{orcidlink}

\begin{document}

\renewcommand{\figureautorefname}{Fig.} %
\renewcommand{\sectionautorefname}{SM.}
\renewcommand{\subsectionautorefname}{Appendix}

\renewcommand{\equationautorefname}{Eq.}
\newcommand{\stepindex}{s} 
\newcommand{\stepid}{^{(\stepindex)}}
\newcommand{\stepprev}{^{(\stepindex-1)}}
\newcommand{\stepmtwo}{^{(\stepindex-2)}}

\title{Quantum-Inspired Simulation of 2D Turbulent Rayleigh-Bénard Convection}

\date{\today}
\author{Nis-Luca van Hülst \orcidlink{0009} \affiliation{Institute for Quantum Physics, University of Hamburg, Luruper Chaussee 149, 22761 Hamburg, Germany}}
\email[Contact author: ]{nis-luca.van.huelst@uni-hamburg.de}

\author{Mario Guillaume Cecile~\texorpdfstring{\orcidlink{0000-0002-2076-6236}}{}}
\affiliation{Institute for Quantum Physics, University of Hamburg, Luruper Chaussee 149, 22761 Hamburg, Germany}

\author{Hai-Yen Van \orcidlink{0009-0002-6301-4293}}
\affiliation{Institute for Quantum Physics, University of Hamburg, Luruper Chaussee 149, 22761 Hamburg, Germany}

\author{Tomohiro Hashizume \orcidlink{0000-0002-7154-5417}}
\affiliation{Institute for Quantum Physics, University of Hamburg, Luruper Chaussee 149, 22761 Hamburg, Germany}
\affiliation{The Hamburg Centre for Ultrafast Imaging, Luruper Chaussee 149, 22761, Hamburg, Germany}

\author{Eugene~de~Villiers \orcidlink{0000-0002-0182-3637}}
\affiliation{
ENGYS Ltd., London SW18 3SX, United Kingdom
}

\author{Dieter Jaksch \orcidlink{0000-0002-9704-3941}}
\affiliation{Institute for Quantum Physics, University of Hamburg, Luruper Chaussee 149, 22761 Hamburg, Germany}
\affiliation{The Hamburg Centre for Ultrafast Imaging, Luruper Chaussee 149, 22761, Hamburg, Germany}
\affiliation{Clarendon Laboratory, University of Oxford, Parks Road, Oxford OX1 3PU, UK}

\begin{abstract}
Turbulent thermal convection governs heat transport in systems 
ranging from stellar interiors to industrial heat exchangers.
Two-dimensional Rayleigh--B\'{e}nard convection serves as a paradigm for these flows, reproducing key features such as thin boundary layers, 
large-scale circulation, and sustained plume dynamics.
While Matrix Product State (MPS) methods have demonstrated significant compression of isothermal turbulent fields, their application to buoyancy-driven flows with active thermal coupling 
has remained unexplored.
We apply MPS to two-dimensional Rayleigh--B\'{e}nard convection
with dynamical simulations up to $\mathrm{Ra} = 10^{10}$.
An \textit{a priori} decomposition of DNS snapshots up to 
$\mathrm{Ra} = 10^{11}$ shows that the bond dimension $\chi$ 
required to represent the flow fields grows without saturation, 
in contrast to the plateauing of~$\chi$ reported for velocity fields 
in isothermal 2D turbulence.
Crucially, however, dynamical simulations solving the governing equations 
directly in the compressed MPS format at fixed~$\chi$ show that the $\chi$ 
required to recover statistical observables, such as the Nusselt number, scales significantly more favorably with $\mathrm{Ra}$ than the \textit{a priori} complexity suggests.
At $\mathrm{Ra} = 10^{10}$, a relative error of $1.8\%$ in the 
mean Nusselt number is achieved with a nearly 9-fold 
reduction in degrees of freedom, using a $\chi$ comparable to that required at $\mathrm{Ra} = 10^{9}$.
Spectral analysis confirms the progressive recovery of spatial and temporal scales with increasing $\chi$. These findings establish MPS as a scalable tool for simulating thermally driven turbulence, suggesting the method may remain viable for investigations of the ultimate regime at substantially higher $\mathrm{Ra}$.
\end{abstract}
\maketitle

\section{Introduction}
\label{sec:intro}
Turbulent thermal convection is ubiquitous in nature, shaping the 
dynamics of planetary and stellar atmospheres~\cite{2011AnRFM..43..583J}, 
oceans~\cite{Kuhlbrodt_2007}, and global climate systems~\cite{Vallis_2006}. The canonical model for such flows is Rayleigh-Bénard convection (RBC), where a fluid layer is heated from below and cooled from above~\cite{Rayleigh1916,Benard1900, Oberbeck1879, Boussinesq1903}. Despite its conceptual simplicity, the nonlinear coupling between velocity and temperature fields gives rise to complex flow phenomena, including 
thermal plumes and large-scale 
circulation~\cite{Bodenschatz2000,RevModPhys.65.851,Ahlers2009}.
The two-dimensional analogue of the system has itself been the subject of extensive study, retaining key aspects of the heat-transport scaling and boundary-layer dynamics of its three-dimensional 
counterpart~\cite{VanDerPoel2015,Zhu2018,Zhang_Zhou_Sun_2017} while 
enabling simulations at substantially higher Rayleigh 
numbers~\cite{Zhu2018}.

A central open question in RBC concerns the existence and nature of the ultimate regime at extreme Rayleigh numbers~\cite{Kraichnan1962,GrossmannLohse2000,GrossmannLohse2011,Ahlers2009, Zhu2018,Shishkina2024, Lohse2024}. In particular, it remains unresolved whether the turbulent heat transport, quantified by the Nusselt number, continues to increase indefinitely with Rayleigh number ($\Ra$), or whether it exhibits a saturation or very weak-growth regime~\cite{Chavanne1997,Ahlers2009}. The primary bottleneck in simulating RBC is the steep increase in computational demand with increasing $\Ra$; the boundary layer thickness decreases rapidly, requiring increasingly fine spatial resolution and small time steps~\cite{Scheel2014}. Addressing this question requires accessing parameter regimes far beyond those currently reachable by conventional numerical techniques, as direct numerical simulations become prohibitively expensive at extreme Rayleigh numbers~\cite{MoinMahesh1998,Pope2000}.

Beyond its physical interest, RBC has established itself as a rigorous 
testbed for state-of-the-art CFD solvers~\cite{Chandrasekhar1961,Busse1978}. Its value as a benchmark lies in its geometric simplicity, which permits discretization on regular Cartesian meshes, together with the absence of empirical turbulence modeling~\cite{MOHANRAI199115}. Resolving the 
nonlinear coupling between the Navier--Stokes equations and an active 
temperature field nonetheless remains computationally demanding, 
especially at high Rayleigh numbers. While parallelization and domain-decomposition strategies can partially mitigate this cost, they typically require substantial computational resources~\cite{VanDerPoel2015,Zhu2018}.

An alternative route is to exploit structural properties of the discretized equations. In particular, uniform Cartesian discretizations give rise to differential operators with an intrinsic low-rank structure, which can be leveraged by tensor-network (TN) representations~\cite{Oseledets2011,Kazeev2012,arenstein2025}. Recent works on two-dimensional turbulent flows have shown that TN representations, in particular matrix product states (MPS)~\cite{White1992,Verstraete2008,Schollwock2011,Orus2014}, can efficiently compress turbulent flow fields while maintaining good accuracy even in strongly nonlinear regimes~\cite{Gourianov2022,Gourianov2022b,Gourianov2024,Hoelscher2024,pisoni2025}. Related progress has also been reported for multi-component reacting shear flows~\cite{pinkston2025}, weakly interacting atomic gases~\cite{connor2025gpe,gomezlozada2025}, and plasma dynamics~\cite{Ye2022,Ameri2023,Ye2024-arxiv,connor2025plasma}. Whether these favorable properties extend to thermally driven flows 
such as RBC, however, remains an open question.

In the present work, we evaluate the practicality of the MPS approach 
for RBC. Building on recent studies of isothermal turbulence in 
periodic domains~\cite{Gourianov2022,Hoelscher2024}, we examine whether 
the favorable compression and computational scaling properties persist 
in this more demanding setting, where buoyancy forcing, thermal 
coupling, and no-slip boundary conditions introduce qualitatively 
new challenges. We report two main findings.
First, an \textit{a priori} analysis of 
turbulent RBC snapshots shows no saturation in the required bond 
dimension~$\chi$ with increasing $\Ra$, indicating that the flow 
complexity, as measured by $\chi$, continues to grow with $\Ra$ in 
the investigated regime. Second, dynamical MPS simulations reveal 
that the $\chi$ required to recover mean heat transport grows 
substantially more slowly than the snapshot analysis would suggest, 
pointing toward a potential scaling advantage over conventional methods.
Since the ``ultimate regime'' demands simulations at Ra 
beyond the reach of current DNS, a favorable scaling of~$\chi$ with~$\Ra$ 
would position MPS as a candidate for its numerical 
investigation~\cite{Zhu2018, GrossmannLohse2000, Ahlers2009}.

The remainder of this paper is organized as follows. Section~\ref{sec:phyiscal_model} outlines the physical model and the numerical schemes for the governing equations, followed by the introduction of the MPS framework in Section~\ref{sec:ten_framework}. Our numerical results are presented in Section~\ref{sec:num_results}, beginning with an \textit{a priori} analysis of turbulent DNS snapshots to determine the effective degrees of freedom (Section~\ref{sec:MPS_turbulent_fields_A}) and a preliminary scaling of $\chi$ with Ra (Section~\ref{sec:scaling_of_complexity_with_Ra_A_priori_B}).
In Section~\ref{sec:temporal_validation_MPS_C}, we perform a preliminary 
validation of the MPS solver by examining the convergence of the simulated 
flow fields and Nusselt number with $\chi$. Section~\ref{sec:turbulent_statistics_D} then presents a comprehensive 
study of the MPS solver at high Rayleigh numbers, 
including an \textit{a posteriori} scaling analysis of the required bond 
dimension with Ra. Finally, 
Section~\ref{sec:discussion} provides 
a discussion of the implications and an outlook for future research.

\begin{figure}[t]
\centering

\includegraphics[width=0.98\columnwidth]{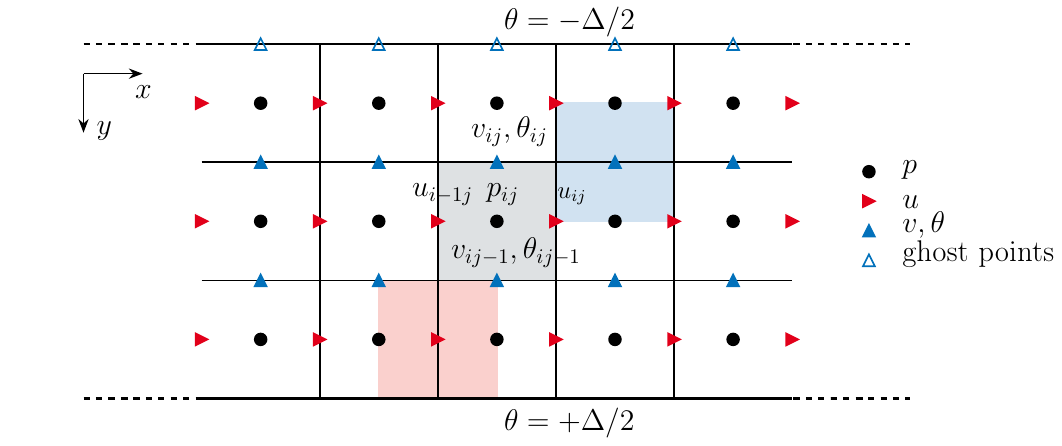}    
\caption{Schematic representation of the staggered grid, showing the locations of pressure $p$, velocity components $u$ and~$v$, temperature $\theta$, and their corresponding ghost points.
}
\label{fig:staggeredgrid}
\end{figure}
\noindent
\section{Physical Model and Numerical Discretization}
\label{sec:phyiscal_model}
In this section, we establish the physical and numerical foundations of our study. We first define the governing equations for thermal convection and subsequently describe the staggered finite difference scheme used for spatial and temporal discretization. Furthermore, it outlines the numerical scheme to solve the equations and introduces the Nusselt number, which will be used throughout this manuscript as a figure of merit.
\subsection{Governing Equations}
The system is described by the incompressible Navier-Stokes equations under the Oberbeck-Boussinesq approximation~\cite{Oberbeck1879, Boussinesq1903, Rayleigh1916}. This approach treats the fluid density $\rho$ as a constant in the inertial and continuity terms, accounting for its variation only in the buoyancy term. The resulting non-dimensional equations for RBC read as~\cite{Chandrasekhar1961, Ahlers2009, VanDerPoel2015}
\begin{align}
\partial_t u_i + u_j \partial_j u_i &= -\partial_i p + \frac{1}{\Re} \partial_j \partial_j u_i + \frac{\Ra}{\Pran \Re^2} \theta \delta_{iy}, \label{eq:momentum} \\ \partial_t \theta + u_j \partial_j \theta &= \frac{1}{\Re \Pran} \partial_j \partial_j \theta, \label{eq:energy} \\ \partial_j u_j &= 0, \label{eq:continuity}
\end{align}
\noindent where repeated indices imply summation following the Einstein notation, $u_i$ denotes the $i$-component ($i\in\{x,y\}$) of the non-dimensional velocity field, namely $u_x=u$ and $u_y=v$, $p$ is the non-dimensional kinematic pressure, $\theta$ is the non-dimensional temperature and $t$ the non-dimensional time. The delta $\delta_{iy}$ indicates that buoyancy acts in the wall-normal direction.
The non-dimensional variables are defined through the Rayleigh number ($\Ra = g \beta \Delta L^3 / \nu \kappa$), representing the buoyancy-to-dissipation ratio, and the Prandtl number ($\Pr = \nu/\kappa$). Following the free-fall scaling convention \cite{Ahlers2009,VanDerPoel2015}, we adopt the cell height $L$, the temperature difference $\Delta$ between the walls, and the free-fall velocity $U_{\ff} = \sqrt{g \beta \Delta L}$ %
as characteristic scales. Here, $g$ is gravity, $\beta$ is the thermal expansion coefficient, $\nu$ is the kinematic viscosity, and $\kappa$ is the thermal diffusivity. This choice yields the characteristic time $t_\ff = L/U_\ff$ %
and relates the Reynolds number to the primary parameters as $\Re = \sqrt{\Ra/\Pr}$.
In the wall-parallel $x$-direction, we assume periodic boundary conditions over a domain of width $H=\Gamma \, L$, where $\Gamma = 2$ is the aspect ratio. At the horizontal plates ($y = \pm L/2$), we enforce no-slip conditions for the velocity, $u = v = 0$, and isothermal Dirichlet conditions for the temperature, $\theta(x, y = \pm L/2) = \pm \Delta/2$ \cite{Ahlers2009, Chandrasekhar1961}.

\subsection{Numerical Scheme}
\label{sec:numericalscheme}
 The state variables are discretized on a staggered grid consisting of $N_x \times N_y$ points per subgrid~\cite{Harlow1965, Ferziger2002:CMFD}. To eliminate interpolation errors in the buoyancy term, $v$ and $\theta$ are co-located on the same subgrid, which is chosen to include the upper boundary. In contrast, the pressure nodes $p$ are located at the cell centers, while the horizontal velocity $u$ is staggered horizontally relative to the pressure (Fig.~\ref{fig:staggeredgrid}) \cite{VanDerPoel2015}. %
 For the operators, we employ a second-order centered finite-difference scheme ~\cite{LeVeque2007}. This low-order approximation is specifically chosen for its robustness in handling the sharp temperature gradients characteristic of high-Rayleigh number convection~\cite{VanDerPoel2015}. Furthermore, from a tensor-network perspective, these low-order operators offer a highly efficient representation of the discrete Laplacian and gradient operators \cite{Kazeev2012}.
 For temporal discretization, we employ a second-order implicit-explicit Runge-Kutta scheme (IMEX-RK2)~\cite{Ascher1997}. In this framework, the advective and buoyancy terms are treated explicitly, while the viscous and thermal diffusion terms are treated semi-implicitly to bypass the stringent quadratic stability constraint ($\delta t \sim \delta x^2$). While higher-order variants, such as the third-order RK3 schemes described by Rai and Moin \cite{MOHANRAI199115} and van der Poel \textit{et al.} \cite{VanDerPoel2015}, are frequently utilized in Rayleigh-Bénard studies for enhanced temporal accuracy, the present work utilizes a robust two-stage IMEX-RK2 approach. This choice provides a balance between computational efficiency and the second-order accuracy required for the flow regimes studied. The incompressibility constraint ($\partial_j u_j = 0$) is enforced via a fractional-step Chorin projection method~\cite{Chorin1968} applied at each stage. Given the fields at time step $m$, the advancement to $m+1$ proceeds as follows:
 \paragraph{Stage 1}A provisional velocity field $u_i^*$ is computed, followed by a projection to obtain the divergence-free field 
 $u_i^{m+1/2}$:
\begin{align}
&u_i^* = u_i^m + \frac{\delta t}{2}\Bigg[
\Big( -u_j^m \partial_j u_i^m 
\nonumber \\ &\qquad \ + \frac{\mathrm{Ra}}{\mathrm{Pr}\,\mathrm{Re}^2}\,\theta^m \delta_{iy} \Big)
 + \frac{1}{\mathrm{Re}} \partial_j \partial_j u_i^*
\Bigg] \nonumber\\
&\partial_j \partial_j \phi^{m+1/2} = \frac{2}{\delta t}\,\partial_j u_j^* \nonumber\\
&u_i^{m+1/2} = u_i^* - \frac{\delta t}{2}\,\partial_i \phi^{m+1/2} \nonumber
\end{align}
 \paragraph{Stage 2}Using the half-step velocity, the final velocity $u_i^{m+1}$ is computed
\begin{align}
&u_i^{*} = u_i^m + \delta t \Bigg[
\Big( -u_j^{m+1/2}\partial_j u_i^{m+1/2}
       \nonumber\\ &\qquad \ \ +\frac{\mathrm{Ra}}{\mathrm{Pr}\,\mathrm{Re}^2}\,\theta^{m+1/2}\delta_{iy} \Big)
+ \frac{1}{\mathrm{Re}}\,\partial_j \partial_j u_i^{m+1/2}
\Bigg] \nonumber \\
&\partial_j \partial_j \phi^{m+1} = \frac{1}{\delta t}\,\partial_j u_j^{*} \nonumber \\
&u_i^{m+1} = u_i^{*} - \delta t\,\partial_i \phi^{m+1} \nonumber
\end{align}
The pressure field is updated as $p^{m+1} = \phi^{m+1}$. The temperature $\theta$ is advanced similarly through the same two stages using the implicit operator $\frac{1}{\text{Re}\,\text{Pr}} \partial_j^2 \theta$ in the first stage. This approach ensures that the time step $\delta t$ is governed by the linear Courant-Friedrichs-Lewy (CFL) condition
$$C = \delta t \left( \frac{|u|}{\delta x} + \frac{|v|}{\delta y} \right) \leq C_{\max}$$ where $C_{\max}$ is the Courant number (typically $C_{\max} < 1$)~\cite{Courant1928}, significantly improving efficiency for high-$\text{Ra}$ simulations.

\subsection{Nusselt Number}
The Nusselt number, denoted as $\mathrm{Nu}$ in the following, serves as our most important system response quantifying the efficiency of convective heat transport relative to pure conduction \cite{Cengel2017}. Physically, it is defined as the ratio of total vertical heat flux to the conductive flux. Below the critical Rayleigh number $\mathrm{Ra}_c$ for the onset of convection, heat transfer is purely diffusive, yielding $\mathrm{Nu} = 1$. As $\mathrm{Ra}$ exceeds $\mathrm{Ra}_c$, fluid 
motion enhances transport, leading to $\mathrm{Nu} > 1$.

We define the instantaneous Nusselt number as
\begin{equation}
\mathrm{Nu}_{\text{inst}}(t) = 1 + \sqrt{\mathrm{Ra}\,\mathrm{Pr}} \, \langle v(t)\,\theta(t) \rangle_V \, ,
\label{eq:Nu_instantaneous}
\end{equation}
where $\langle \cdot \rangle_V$ denotes the spatial average over the domain. To characterize the statistically stationary heat transport, we consider the time-averaged Nusselt number
\begin{equation}
\mathrm{Nu}(\DT) = \frac{1}{\DT} \int_{t_0}^{t_0 + \DT} \mathrm{Nu}_{\text{inst}}(t)\, dt \, ,
\label{eq:Nu_time_avg}
\end{equation}
where $\DT$ is the averaging window and $t_0$ is chosen to exclude initial transients.

Empirically, the Nusselt number follows a power-law scaling with the Rayleigh number, $\mathrm{Nu} \sim \mathrm{Ra}^{\gamma}$, where the exponent $\gamma$ transitions from $\approx 1/4$ in the moderate regime to $\approx 1/3$ in the fully turbulent regime \cite{Ahlers2009,GrossmannLohse2000}.

Furthermore, the distribution of instantaneous Nusselt-number fluctuations,
\(\mathrm{Nu}_{\mathrm{inst}}(t)\), serves as a robust metric for assessing the numerical solver, and we use it here to benchmark the tensor-network-based simulations against a DNS reference. To quantify the discrepancy between two such distributions, we employ
the first Wasserstein distance \(\tilde{W}_1\)~\cite{Vallender1974, Givens1984}. For distributions with
cumulative distribution functions (CDFs) \(U(x)\) and \(V(x)\), it is defined as the
\(\mathrm{L}_1\) distance between the CDFs~\cite{Vallender1974}
\begin{equation}
    \label{eq:W1}
    \tilde{W}_1 = \int_{-\infty}^{\infty} |U(x) - V(x)| \, dx \,.
\end{equation}

To facilitate scale-independent comparisons across varying thermal driving regimes, we introduce the normalized Wasserstein distance, $W_1 = \tilde{W}_1 / \mathrm{Nu}$, which scales the distribution error relative to the mean macroscopic heat transport.

\section{Tensor Network Framework}
\label{sec:ten_framework}
To mitigate the memory and computational demands of high-$\Ra$ simulations, we adopt a tensor-network representation. Instead of storing discretized fields as conventional multidimensional arrays, each field is encoded as a Matrix Product State (MPS). We consider a two-dimensional discretized field $f(x_{i_x}, y_{i_y})$ sampled on a uniform grid. The mapping from the integer index pair $(i_x, i_y)$ to the physical domain is given by
\begin{equation}
(x, y) = (x_0, y_0) + (i_x \Delta x, i_y \Delta y) \nonumber
\end{equation}
where $(x_0, y_0)$ is the origin and $\Delta x$, $\Delta y$ are the grid spacings, and $i_x \in \{0, \dots, 2^{n_x}-1\}$, $i_y \in \{0, \dots, 2^{n_y}-1\}$. Each integer index can be expressed as a binary string of length $n_x$ or $n_y$, respectively, as $i_x = \sum_{k=1}^{n_x} s_k 2^{n_x-k}$, where $s_k \in \{0,1\}$. By concatenating these bitstrings, the 2D field is reinterpreted as a rank-$n$ tensor $f(s_1, \dots, s_n)$ with $n = n_x + n_y$. This tensor can then be factorized via a sequence of singular value decompositions (SVDs) into a chain of $n$ third-order tensors, or MPS cores $A^{(l)}$. Under this ansatz, the field value at any grid location is recovered by contracting these cores along their virtual indices $\alpha$
\cite{Schollwock2011, Vidal2003}
\begin{align}
    f(s_1, ..\,, s_n) &\simeq \sum_{\alpha_1=1}^{\chi_1} ..\, \sum_{\alpha_{n-1}=1}^{\chi_{n-1}}  A_{s_1}^{(1) \alpha_1} A_{s_2}^{(2) \alpha_1 \alpha_2} \dots A_{s_n}^{(n) \alpha_{n-1}} \nonumber \\
    &=
    \begin{tikzpicture}[tensornetwork, scale=1.0, xscale=0.9, baseline=(A1.base)]
        \node[atensor] (A1) at (1, 0) {$A^{(1)}$};
        \node[atensor] (A2) at (2.2, 0) {$A^{(2)}$};
        \node (dots) at (3.2, 0) {$\dots$};
        \node[atensor] (An) at (4.2, 0) {$A^{(n)}$};
        \draw (A1.north) -- +(0, 0.4) node[above] {$s_1$};
        \draw (A2.north) -- +(0, 0.4) node[above] {$s_2$};
        \draw (An.north) -- +(0, 0.4) node[above] {$s_n$};
        \draw (A1) -- node[above, font=\scriptsize] {$\alpha_1$} (A2);
        \draw (A2) -- node[above, font=\scriptsize] {$\alpha_2$} (dots);
        \draw (dots) -- node[above, font=\scriptsize] {$\alpha_{n-1}$} (An);
    \end{tikzpicture} \nonumber.
\end{align}

The bond dimension $\chi=\max(\chi_{l})$ determines the representational capacity of the MPS. For fields exhibiting low-rank entanglement, this representation provides a compressed format where storage costs scale logarithmically, $O(\log{N} \, \chi^2)$, with the number of grid points $N$~\cite{vanhülst2025,vanhülst2025qsolvers}.
 
\begin{figure*}[t]
    \centering %
    \includegraphics[width=\textwidth]{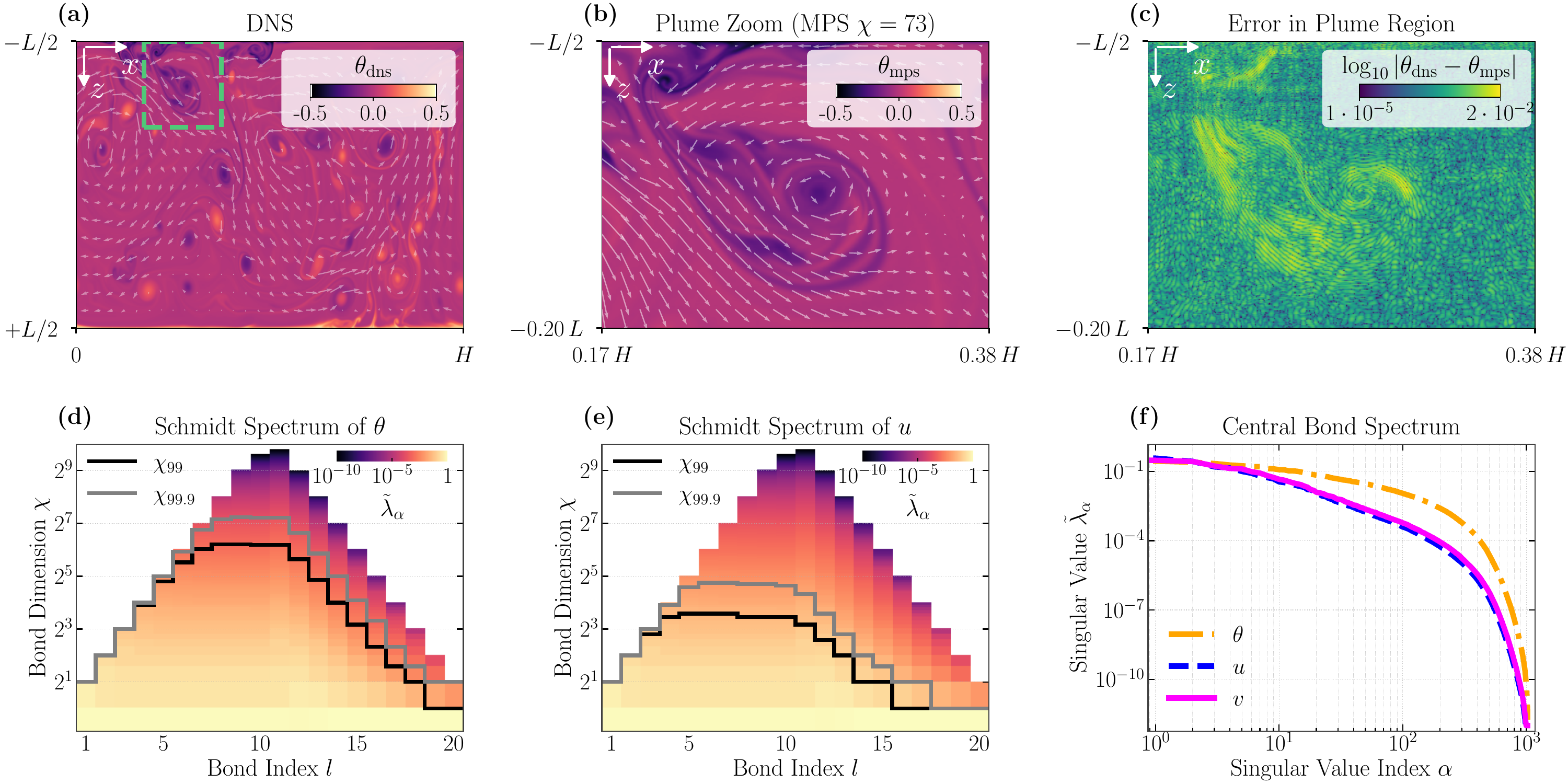}
    \caption{
    \textbf{A priori MPS compression and entanglement structure of turbulent RBC fields at $\boldsymbol{\Ra=10^{10}}$.} All panels display data from a reference DNS on a $2^{11}\times 2^{10}$ grid (full simulation parameters are detailed in Appendix \ref{sec:app:apriori_scaling}).
    (a) Instantaneous temperature field $\theta$ at $\Ra=10^{10}$ from a DNS simulation on a $2^{11}\times 2^{10}$ grid.
    (b) Compressed MPS reconstruction of the same field (focusing on the green square in panel (a)), retaining only $20\%$ of the original parameters ($P_T = 0.2$). The visual features of thermal plumes are preserved qualitatively.
    (c) Pointwise error field $|\theta_{\text{DNS}} - \theta_{\text{MPS}}|$, illustrating that compression errors are localized to the regions around thermal plumes.
    (d--e) Spatial distribution of the Schmidt spectrum (entanglement entropy) along the MPS for the temperature field $\theta$ (d) and horizontal velocity $u$ (e). The heatmaps display the magnitude of singular values at each bond. Overlaid lines indicate the local bond dimension $r$ required to retain $99\%$ (solid) and $99.9\%$ (dashed) of the spectral weight (see Eq.~\eqref{eq:truncation_criterion}). %
    (f) Decay of the normalized singular values $\tilde{\lambda}_\alpha$ (Schmidt coefficients) plotted against the index $\alpha$ at the central bipartition for $u, v,$ and $\theta$. %
    }
    \label{fig:compression}
\end{figure*}

To assess the attainable compression of the turbulent flow fields obtained by DNS, we analyze the Schmidt spectrum of the resulting MPS upon decomposition. We define a truncation criterion based on the cumulative spectral weight, calculating the minimum bond dimension $r$ required at each bond to retain a fraction $1-\epsilon$ of the total squared weight,
\begin{equation}
    \frac{\sum_{\alpha=1}^{r} \lambda_{\alpha}^2}{\sum_{\alpha=1}^{\chi} \lambda_{\alpha}^2} \ge 1 - \epsilon,
    \label{eq:truncation_criterion}
\end{equation}
where $\lambda_{\alpha}$ are the Schmidt coefficients at a given bond, sorted in descending order.
This metric allows us to map the information content of the flow, for example, a tolerance of $\epsilon = 10^{-2}$ corresponds to retaining $99\%$ of the spectral weight at each bond, denoted as $\chi_{99}$~\cite{Gourianov2022}. Finally, we introduce the fraction in the number of parameters $P_T$ of the MPS compared to the DNS \cite{pinkston2025}
\begin{align}
\label{eq:Pt}
    P_T = \frac{\sum_{l=1}^{n} \chi_{l-1} \, d \, \chi_l}{2^n} \ ,
\end{align}
where $d=2$ is the physical dimension in our simulations. This figure of merit will be used throughout the results as a measure of the parametric efficiency of the MPS.

For the representation of operators and the implementation of algebraic operations in the MPS format, including operator application, element-wise state multiplication, and the solution of linear systems, we refer the reader to our recent work~\cite{vanhülst2025}. The momentum equations are solved using a single-site density matrix renormalization group approach \cite{Schollwock2011, Holtz2012, Dolgov2012fastmatvec, Dolgov2013AlternatingME}, also known as the alternating minimization scheme, employing Krylov-based methods for the occurring local linear systems. The pressure Poisson equation is solved via a spectral method adapted for the MPS format, inspired by~\cite{Chen2023}. The required Fourier transform operators in MPS format are generated using the library introduced in~\cite{Nunez2025}.

\section{Numerical Results}
\label{sec:num_results}

\subsection{Matrix Product State Representation of Turbulent Flow Fields} 
\label{sec:MPS_turbulent_fields_A}
This section examines the data compression capability of the MPS format for turbulent RBC flow fields. Snapshots of the velocity components $u$, $v$ and temperature $\theta$ are taken from DNS solutions of Eqs.~\eqref{eq:momentum}--\eqref{eq:continuity}, evolved from the conductive state to statistical stationarity on the staggered grid
introduced in Fig.~\ref{fig:staggeredgrid}. The grid resolution is taken according to Zhu \textit{et al.}~\cite{Zhu2018}, and time integration is performed using the RK2 scheme detailed in Sec.~\ref{sec:numericalscheme}. Simulation parameters and initialization details are provided in Appendix~\ref{sec:app:apriori_scaling}.
Figure~\ref{fig:compression}(a) displays a representative DNS snapshot of the instantaneous velocity components $u$, $v$ and temperature $\theta$ at $\Ra = 10^{10}$, with the velocity overlaid as a quiver plot. The flow exhibits the characteristic phenomenology of high-Ra convection, featuring thin thermal boundary layers and the detachment of thermal plumes that drive the heat transport across the bulk. To quantify the potential compression of this turbulent state, we decompose the DNS flow fields into MPS via successive SVDs (as detailed in Sec.~\ref{sec:ten_framework}). We truncate the MPS such that the compressed state retains only $20\%$ of the original parameters ($P_T = 0.2$). The fidelity of this compression is illustrated in Fig.~\ref{fig:compression}(b), which shows a detailed zoom of a rising thermal plume reconstructed from the truncated MPS. The pointwise error, $|\theta_{\text{DNS}} - \theta_{\text{MPS}}|$, within this zoomed-in region is shown in Fig.~\ref{fig:compression}(c). This figure validates the accuracy of the reconstruction, with absolute errors remaining below $2 \cdot 10^{-2}$. %

To understand the characteristics of the MPS representation in more depth, we analyse the normalized 
Schmidt spectrum across every bond of the MPS. 
We define the normalized Schmidt values 
$\tilde{\lambda}_{\alpha} = \lambda_\alpha/\sum_{k}\lambda_{k}^2$, 
where the accumulated sum of the largest $\tilde{\lambda}_{\alpha}$ defines the strength of the 
correlations between different length scales \cite{Gourianov2022}. 
In Figures~\ref{fig:compression}(d) and (e) we present the normalized Schmidt values at each bond as a heatmap
for the horizontal velocity $u$ and the temperature $\theta$, respectively.
Overlaid on these heatmaps are the $\chi_{99}$ (solid) and $\chi_{99.9}$ (dashed) lines, 
which indicate the $\chi$ required to retain $99\%$ and $99.9\%$ of the singular value weight (Eq.~\eqref{eq:truncation_criterion}), respectively.

Finally, to examine the convergence rate, we plot the full Schmidt spectrum at the central bipartition for $u, v$, and $\theta$ in Fig.~\ref{fig:compression}(f). The velocity components $u$ and $v$ display nearly identical spectral decay. The temperature spectrum is notably flatter at low indices, confirming the higher information content observed in the heatmaps. However, for all fields, the spectrum transitions into a rapid exponential decay after a critical bond index (around $\alpha \approx 300$). This exponential tail is the crucial feature that supports the tensor network approach. It signifies that the error of the MPS approximation converges exponentially with $\chi$, guaranteeing that simulations become exponentially more accurate as $\chi$ grows.

\subsection{A Priori Scaling of Bond Dimension $\chi$ with Rayleigh Number}
\label{sec:scaling_of_complexity_with_Ra_A_priori_B}
The scaling of the required $\chi$ with increasing flow complexity, parameterized here by Ra, serves as an important indicator for the practicality of the MPS representation in the specific problem setting. We evaluate this scaling behavior by performing reference DNS simulations across the range $10^8 \le \mathrm{Ra} \le 10^{11}$. For each configuration, the system is evolved to a statistically stationary state before representative flow snapshots are extracted for analysis. Further details on the 
equilibration times and grid parameters are provided in Appendix~\ref{sec:app:apriori_scaling}. We note that this \textit{a priori} decomposition of DNS snapshots constitutes a preliminary assessment. While a rigorous determination of scaling laws requires dynamical MPS simulations, this approach provides an initial measure of the expected cost associated with increasing Ra.

\begin{figure}[t]
    \centering
    \includegraphics[width=0.98\columnwidth,trim={0 0 0 0},clip]{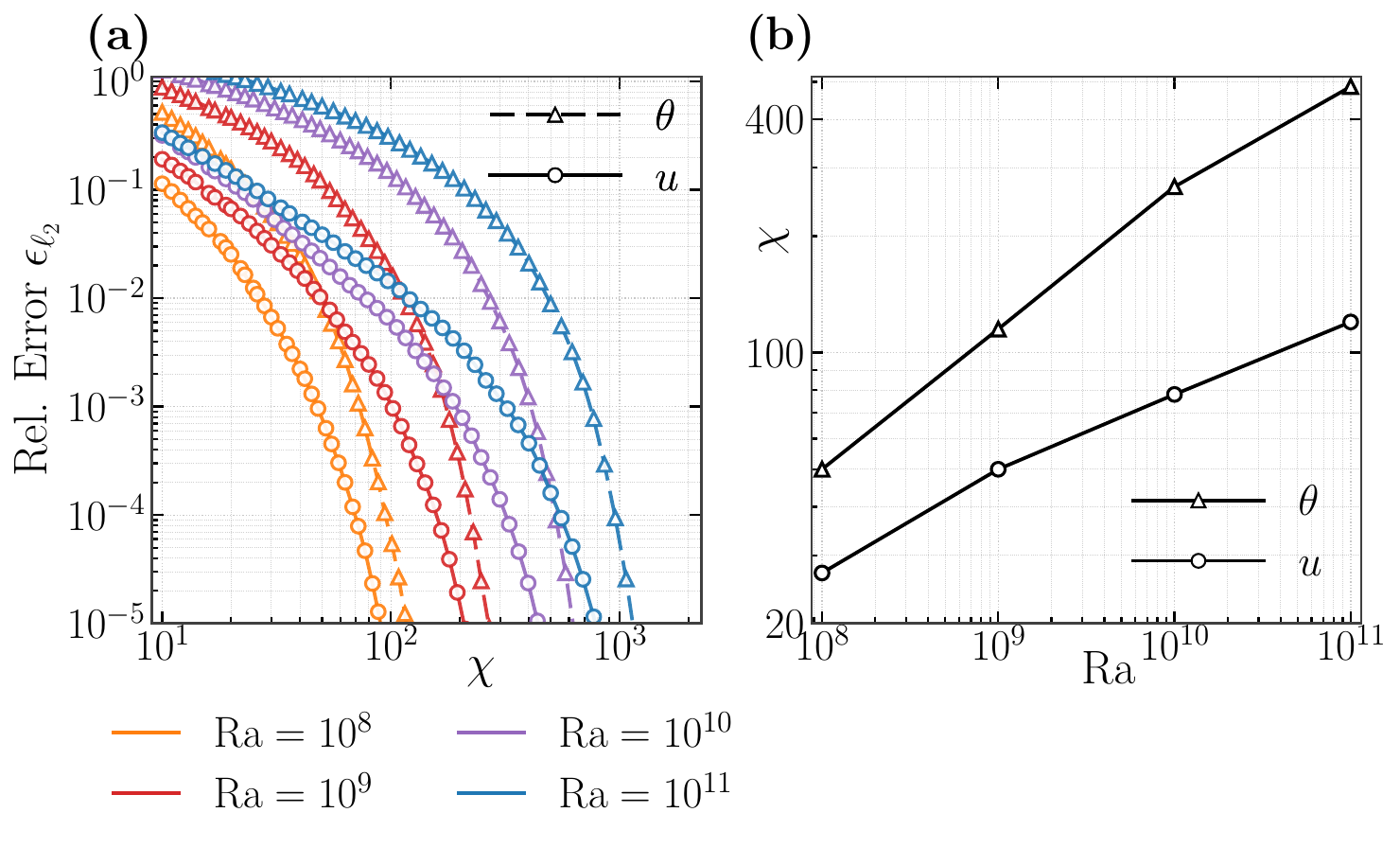}
    \caption{
\textbf{MPS compression accuracy and bond dimension scaling for turbulent RBC fields.}
(a) Relative $\ell_2$-error versus $\chi$ for various $\Ra$, shown for $u$ and $\theta$. Results for $v$ are nearly identical to $u$ (see Appendix~\ref{sec:app:v_scaling}).
(b) Scaling of the minimum $\chi$ required to achieve $\epsilon_{\ell_2} < 10^{-2}$ with $\Ra$, interpolated from the convergence data in~(a).
}
\label{fig:scaling}
\end{figure}

Fig.~\ref{fig:scaling}(a) shows the convergence of the relative $\ell_2$-error,
\begin{equation}
    \epsilon_{\ell_2} = \frac{\|f^{\mathrm{MPS}} - f^{\mathrm{DNS}}\|_2}{\|f^{\mathrm{DNS}}\|_2},
    \label{eq:l2error}
\end{equation}
as a function of $\chi$ for both $u$ and $\theta$ across various Ra, where $f$ denotes the respective field. In Fig.~\ref{fig:scaling}(b), we show the \(\chi\) required to maintain a fixed error tolerance as $\Ra$ increases.
A related analysis was reported by Hölscher \textit{et al.}~\cite{Hoelscher2024} 
for isothermal 2D turbulent flows (a decaying jet and decaying 
homogeneous turbulence), where the required $\chi$ for the velocity fields was observed to saturate at higher Reynolds numbers.
In our case, the growth of $\chi$ for $u$ weakens at higher $\mathrm{Ra}$, 
which may indicate the onset of a similar trend, though a clear plateau 
is not yet visible.
For $\theta$, the required $\chi$ continues 
to grow approximately as a power law over the investigated range.
This suggests that $\theta$ is harder to compress in the MPS format than $u$, indicating a strong sensitivity of the thermal field to fluctuations and the resulting increase in flow complexity. %

\begin{figure}[t]
    \centering
    \includegraphics[width=0.98\columnwidth,trim={0 0 0 0},clip]{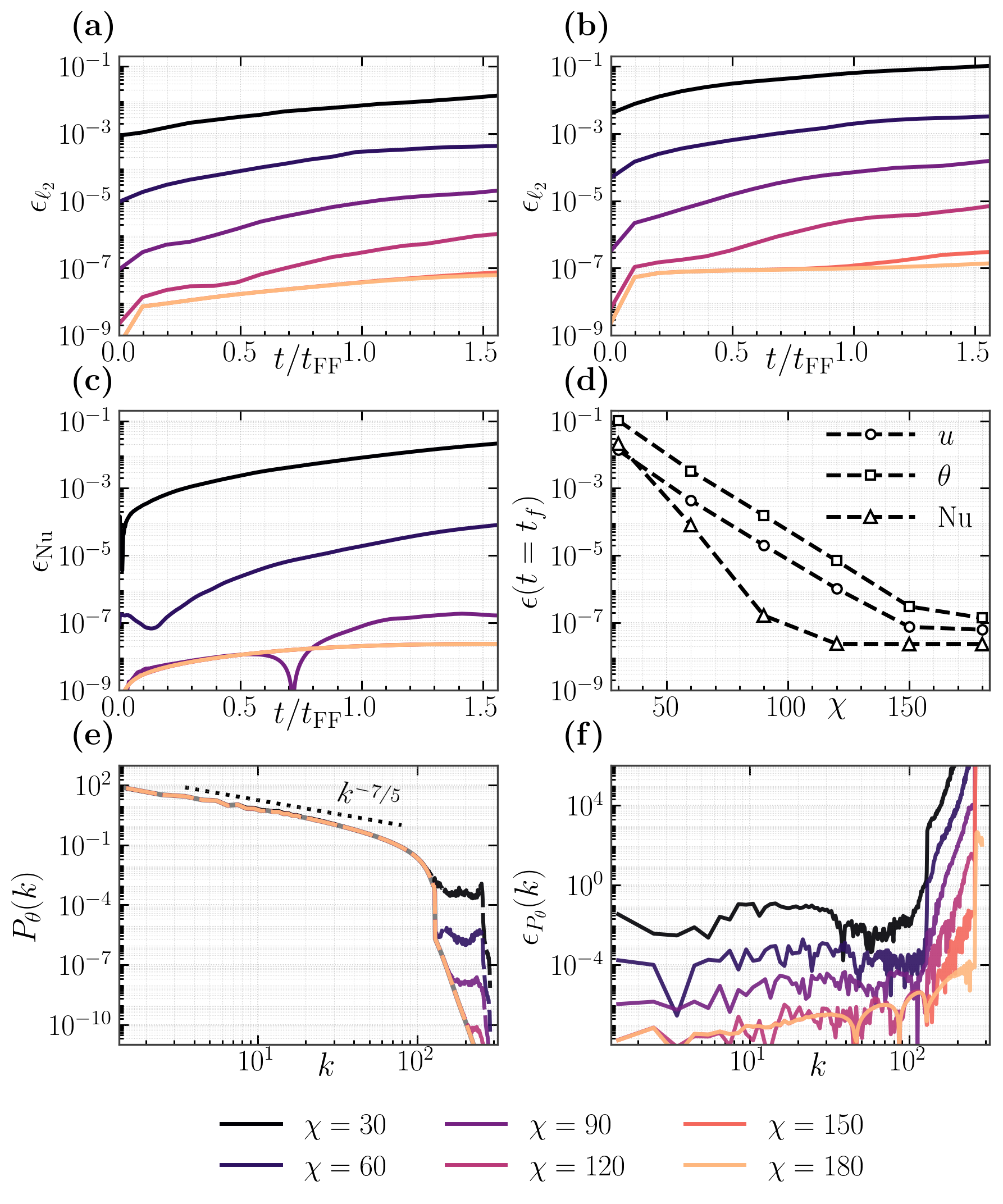}
\caption{\textbf{Dynamical validation at $\boldsymbol{\Ra=10^{8}}$.}
(a)--(c)~Time evolution of $\epsilon_{\ell_2}$ (Eq.~\ref{eq:l2error}) for 
(a)~velocity~$u$, (b)~temperature~$\theta$, 
and (c)~$\epsilon_{\mathrm{Nu}}$.
(d)~Convergence of $\epsilon$ at $t_f = 1.56\,t_\mathrm{FF}$ 
as a function of $\chi$.
(e)~Temperature power spectral density $P_\theta(k)$ at 
$t = 1.56\,t_\mathrm{FF}$. The dashed line indicates the classical Bolgiano--Obukhov scaling $k^{-7/5}$ for the temperature spectrum~\cite{Bolgiano1959,Samuel2024}.
(f)~Pointwise relative spectral error 
$\epsilon_{P_\theta}(k) 
= |P_\theta^\mps(k) - P_\theta^{\mathrm{DNS}}(k)| 
/ P_\theta^{\mathrm{DNS}}(k)$.}
\label{fig:dynamics_verification}
\end{figure}

\subsection{Accuracy and Convergence of the MPS Solver in Dynamical Simulations}
\label{sec:temporal_validation_MPS_C} 
While the \textit{a priori} decomposition of DNS snapshots provides an initial measure of the informational content of the flow, it tests representability only at discrete points in time.
In the dynamical simulations, the system is evolved directly in the 
compressed MPS format at fixed $\chi$. It is possible that the high $\chi$ suggested by the DNS data reflect small-scale features or fluctuations that have limited influence on the macroscopic evolution of the flow. Consequently, a substantially smaller $\chi$ may suffice to propagate the flow in time while faithfully predicting statistical observables of interest. 
To assess this, we validate the MPS solver by comparing against a DNS reference 
at $\Ra = 10^8$, a regime of unsteady convection where the flow is chaotic but exhibits a sufficiently long Lyapunov~\cite{Vastano1991-lyapunov} time 
to permit direct pointwise comparison over relevant timescales.
The relative $\ell_2$-error $\epsilon_{\ell_2}(t)$ (Eq.~\ref{eq:l2error}) for $u$ and $\theta$ 
are shown in Fig.~\ref{fig:dynamics_verification}(a)--(b). 
Panel~(c) shows the relative error in $\mathrm{Nu}$, 
$\epsilon_{\mathrm{Nu}}(t) = |\mathrm{Nu}^{\mathrm{MPS}}(t) - \mathrm{Nu}^{\mathrm{DNS}}(t)| / \mathrm{Nu}^{\mathrm{DNS}}(t)$, 
where $\mathrm{Nu}(t)$ is averaged up to time $t$.
As illustrated in panels~(a)--(c), the MPS solution converges systematically 
toward the DNS reference as $\chi$ increases.
Panel~(d) summarizes this convergence at $t_f = 1.56\,t_\mathrm{FF}$.
The Nusselt number error $\epsilon_{\mathrm{Nu}}$ decreases substantially 
faster than the field errors $\epsilon_u$ and $\epsilon_\theta$.
This is expected because $\mathrm{Nu}$ is a spatio-temporally averaged observable in which local errors partially cancel, 
whereas the $\ell_2$~errors are of discrete snapshots and retain sensitivity to deviations 
across all scales.

\begin{figure}[t]
    \centering
    \newlength{\topfigwidth}
    \setlength{\topfigwidth}{0.96\columnwidth}

    \begin{tikzpicture}
        \node[anchor=south west, inner sep=0] (topimage) at (0,0) {
            \includegraphics[width=\topfigwidth]{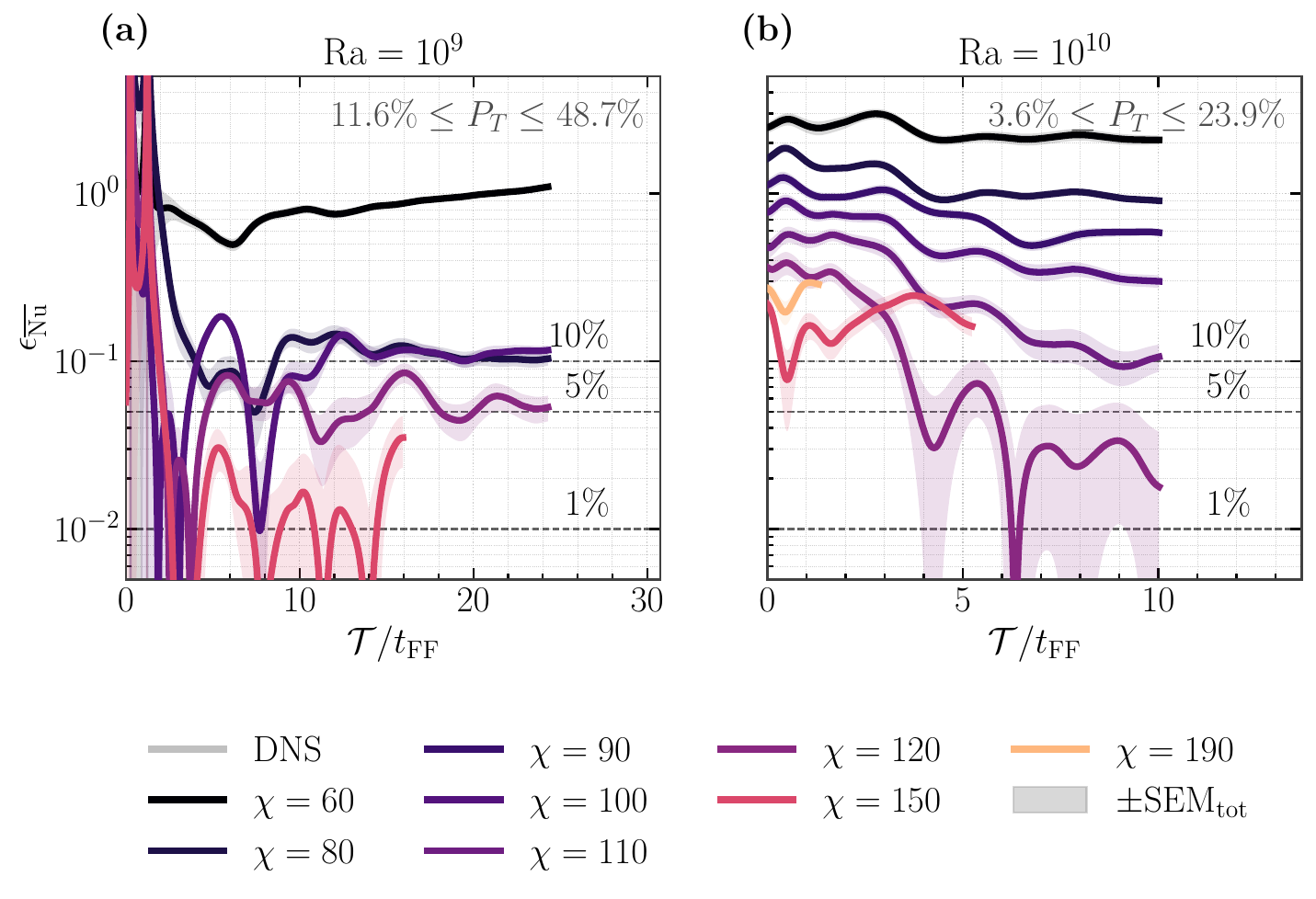}
        };
    \end{tikzpicture}
    \vspace{0.2cm}

    \begin{tikzpicture}
        \node[anchor=south west, inner sep=0] (topimage) at (0,0) {
            \includegraphics[width=\topfigwidth]{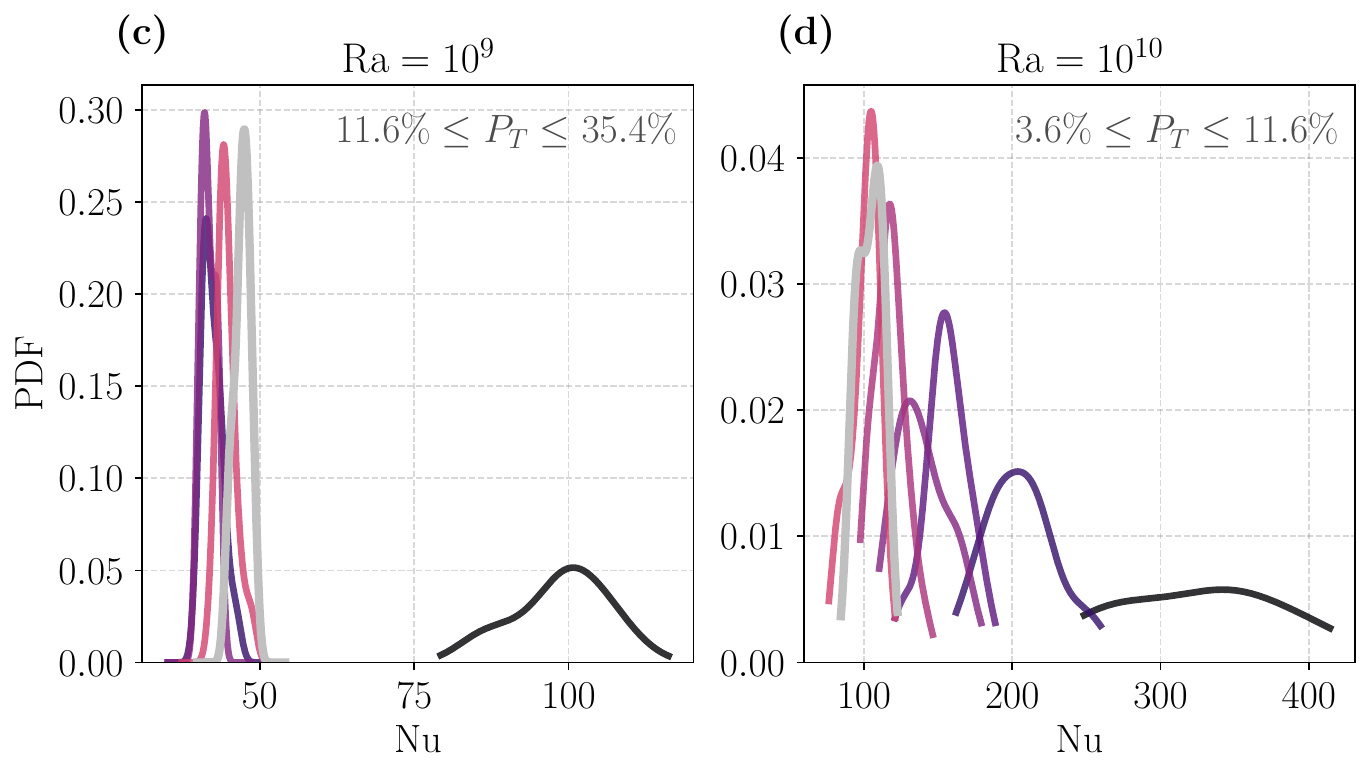}
        };
    \end{tikzpicture}
    \caption{
    \textbf{Convergence of turbulent heat transport.} (a–b) Evolution of the ensemble-averaged relative error $\epsilon_{\Nuens}(\mathcal{T})$ with increasing sampling window $\mathcal{T}$ for (a) $\mathrm{Ra}=10^9$ ($M=19$) and (b) $\mathrm{Ra}=10^{10}$ ($M=24$). Horizontal dashed lines represent $10\%$, $5\%$, and $1\%$ error thresholds. Shaded bands indicate the standard error of the mean 
(SEM, cf.\ Appendix~\ref{sec:app:Nu_error_prop}). (c–d) Probability density functions (PDFs) of the spatio-temporally averaged Nusselt numbers $\Nu_m$ from the $M$ individual realizations at the maximum sampling window ($\mathcal{T} = 24.3\, t_{\ff}$ and $\DT = 10.0\, t_{\ff}$, respectively) for (c) $\text{Ra}=10^9$ and (d) $\text{Ra}=10^{10}$. The range of parameter fractions $P_T$ corresponding to the employed $\chi$ is indicated in the upper right of each panel. Furthermore, more detailed $P_T$ values are provided in Appendix~\ref{sec:app:data_compression}.
    }

    \label{fig:Nusselt_mean}
\end{figure}

To complement these error measures, the temperature spectrum reveals how MPS compression affects the scale-resolved structure of the solution. It is defined as the total spectral power of the Fourier modes lying within the wavenumber shell of radius~$k$~\cite{Samuel2024},
\begin{equation}
    P_\theta(k) = \frac{1}{2\pi} 
    \sum_{|\mathbf{k}'| \in [k,\, k+\mathrm{d}k)} 
    |\hat{\theta} (\mathbf{k}')|^2 .
\end{equation}
Figure~\ref{fig:dynamics_verification}(e) shows $P_\theta(k)$ at $t = 1.56\,t_\ff$ for the 
DNS and MPS solutions at various~$\chi$, alongside the classical 
Bolgiano--Obukhov reference slope $k^{-7/5}$~\cite{Bolgiano1959,Samuel2024}.
The corresponding relative spectral error $\epsilon_{P_\theta}(k)$ in Fig.~\ref{fig:dynamics_verification}(f) exhibits two distinct behaviors.
At low and intermediate wavenumbers, $\epsilon_{P_\theta}$ 
is approximately independent of~$k$ 
and shifts downward uniformly with increasing~$\chi$.
At high wavenumbers, $\epsilon_{P_\theta}$ grows rapidly with~$k$ 
as the DNS spectrum decays exponentially while the MPS spectrum plateaus.
At the present $\mathrm{Ra}$, this discrepancy does not appear to 
affect the convergence of integrated observables such as~$\mathrm{Nu}$.

\subsection{Turbulent Heat Transport at High Rayleigh Numbers}
\label{sec:turbulent_statistics_D}

\begin{figure}[t]
    \centering
    \includegraphics[width=0.9\columnwidth]{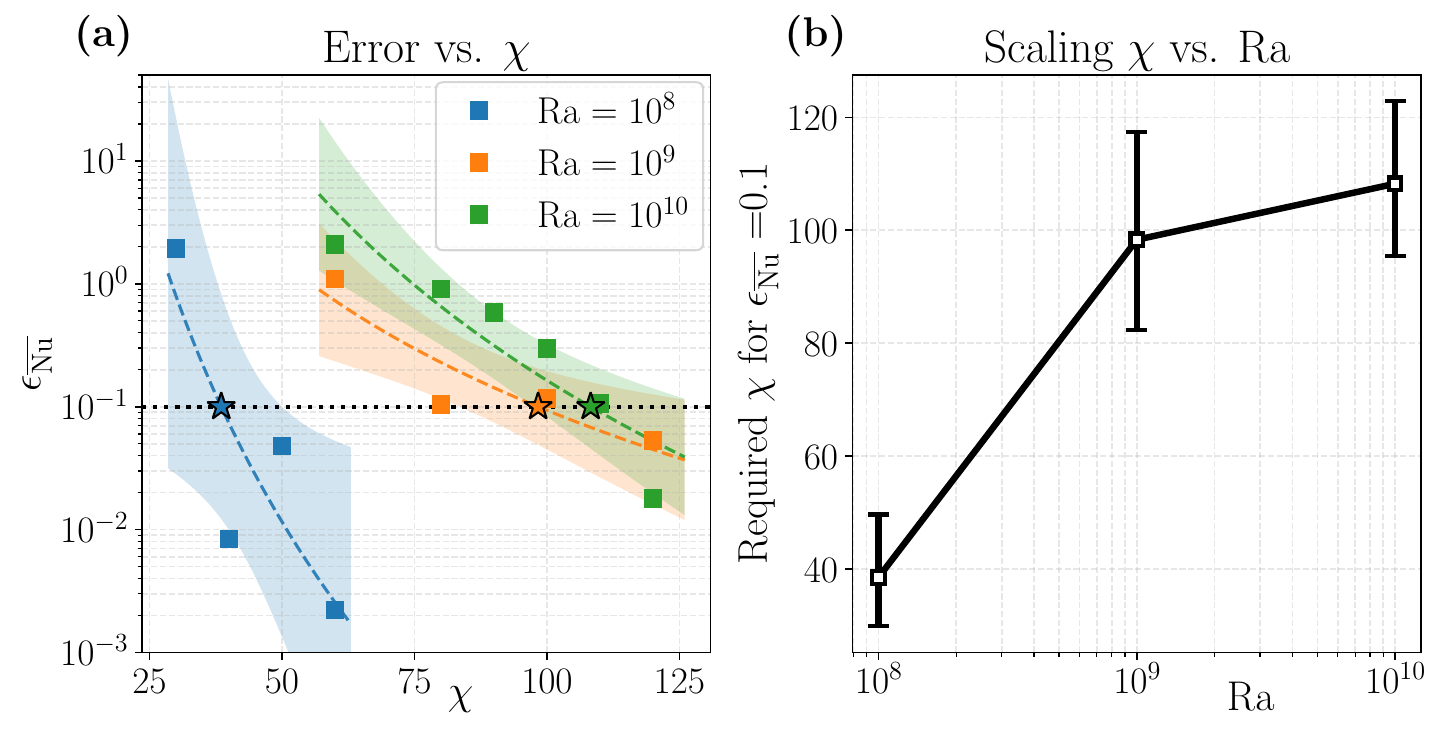}
    \caption{\textbf{Scaling analysis.}
(a) Relative error in the ensemble-averaged Nusselt number,
$\epsilon_{\Nuens}$,
shown as a function of $\chi$ for $\mathrm{Ra}=10^{8},10^{9},10^{10}$.
For each $\chi$, $\overline{\mathrm{Nu}}$ is computed from $M$ realizations at the maximum sampling window $\DT$, for $\mathrm{Ra}=10^{9}$ and $\mathrm{Ra}=10^{10}$ the same $M, \DT$ values as in the corresponding Fig.~\ref{fig:Nusselt_mean}, while $\DT=25.6$ for $\mathrm{Ra}=10^{8}$. 
Dashed lines show power-law fits.
Shaded regions indicate the associated $95\%$ confidence band for the fitted mean trend. %
Stars mark the $\chi$ value at which the fitted trend intersects the target error $\epsilon_{\Nuens}=0.1$
(horizontal dotted line).
(b) Required $\chi$ to reach the target error $\epsilon_{\Nuens}=0.1$ as a function of $\mathrm{Ra}$,
obtained from the fitted trends in panel (a).
}
    \label{fig:Scaling_Chi_vs_Ra}
\end{figure}

At higher Rayleigh numbers, $\text{Ra}=10^9$ and $10^{10}$, the chaotic nature of the flow renders individual trajectory tracking unsuitable, as infinitesimal perturbations lead to rapid divergence of initially close solution trajectories. We therefore assess the MPS approach by its ability to reproduce long-term statistics rather than individual paths. Assuming the ergodic hypothesis \cite{Zhang_Zhou_Sun_2017,Xu2020}, where the time average of a single realization is equivalent to the ensemble average over independent trajectories, we adopt a parallelized sampling strategy to circumvent the prohibitive cost of computing a single long-duration trajectory. We construct an ensemble of $M$ independent realizations, treating them as statistically uncorrelated samples from the attractor.\\
\indent To construct this ensemble, all realizations are initialized from a common statistical stationary state obtained via DNS, with distinct thermal perturbations $\theta_{\text{pert}}$ introduced to trigger trajectory divergence (see Appendix~\ref{app:sec:perturbations}). A slightly higher perturbation intensity is prescribed for the DNS reference to match the inherent truncation errors of the MPS representation, which can be interpreted as a persistent noise source.
Following a transient decorrelation period, we compute the ensemble-averaged Nusselt number $\Nuens = 1/M \, \sum_{m=1}^M \Nu_m$, where $\Nu_m$ is the Nusselt number of the $m$-th realization, and define the relative error $\epsilon_{\Nuens}(\mathcal{T}) = | \Nuens^\mps(\mathcal{T}) - \Nuens^\dns(\mathcal{T}) | / \Nuens^\dns(\mathcal{T})$ for a given sampling window $\DT$. The results for ensembles of $M=19$ realizations at $\text{Ra}=10^9$ and $M=24$ realizations at $\text{Ra}=10^{10}$ are summarized in Fig.~\ref{fig:Nusselt_mean}. Panels (a) and (b) display the evolution of the relative error $\epsilon_{\Nuens}(\mathcal{T})$ over increasing sampling windows $\DT$. In addition, Fig.~\ref{fig:Nusselt_mean}(c) and (d) show the probability density functions (PDFs) of the spatio-temporally-averaged Nusselt numbers $\text{Nu}_m(\DT)$ of the individual realizations based on the maximum sampling window ($\mathcal{T} = 24.3\, t_{\ff}$ and $10.0\, t_{\ff}$, respectively). %

For the $\Ra=10^9$ case (a), the behavior of $\epsilon_{\Nuens}(\DT)$ indicates an abrupt transition in the accuracy of the results rather than a smooth improvement. There is a stark contrast between the results at the lowest $\chi$ and all subsequent increments. At $\chi=60$ ($P_T \approx 11.6\%$), the MPS proves insufficient to recover the heat transport, with $\epsilon_{\Nuens}(\DT)$ saturating near $100\%$ and even increasing weakly as the sampling window grows. The corresponding distribution in Fig.~\ref{fig:Nusselt_mean}(c) confirms this systematic bias, showing a PDF that is entirely detached from the DNS reference and shifted toward significantly higher Nusselt numbers. An increment to $\chi=80$ ($P_T \approx 18.2\%$) yields a substantial improvement, with $\epsilon_{\Nuens}(\DT)$ dropping to the $10\%$ regime. 
Further increases to $\chi=100$ and $\chi=120$ ($P_T \le 35.4\%$) provide only incremental improvements, with errors settling within the $5\%$--$10\%$ and $1\%$--$5\%$ ranges, respectively. This behavior suggests that a minimum parameter fraction between $11\%$ and $18\%$ is necessary at this Ra to faithfully resolve the convective heat transport. %

At the higher Rayleigh number ($\text{Ra}=10^{10}$), the error evolution in Fig.~\ref{fig:Nusselt_mean}(b) exhibits a comparable threshold behavior, though the deviation from the DNS reference extends to higher $\chi$. In this regime, both $\chi=60$ and $\chi=80$ prove insufficient to recover the heat transport, with $\epsilon_{\Nuens}(\mathcal{T})$ exceeding $100\%$. This is also reflected in Fig.~\ref{fig:Nusselt_mean}(d), where the distributions for $\chi$ do not yet overlap with the reference PDF. %
However, an increment to $\chi=100$ results in a transition toward the DNS reference, reducing the error to approximately $30\%$. 
Further refinement to $\chi=120$ brings $\epsilon_{\Nuens}(\mathcal{T})$ consistently below the $10\%$ threshold as the sampling window grows, a result supported by the PDF in Fig.~\ref{fig:Nusselt_mean}(d), which shows significant overlap with the reference distribution. Notably, for $\text{Ra}=10^{10}$, $\chi=120$ corresponds to a parameter fraction of only $P_T \approx 11.6\%$, representing a nearly nine-fold compression.

This threshold is substantially lower than the $35.4\%$ fraction required for comparable accuracy at $\Ra=10^9$, indicating an improved compression efficiency of the MPS representation at higher $\Ra$. Motivated by this observation, Fig.~\ref{fig:Scaling_Chi_vs_Ra} summarizes a qualitative scaling analysis. Panel~(a) shows the relative error $\epsilon_{\Nuens}$ as a function of $\chi$ for $\Ra=10^{8},10^{9},10^{10}$ and (b) the corresponding power-law fits. We determine the $\chi$ for which the fitted trend falls below the prescribed target error $\epsilon_{\Nuens}=10^{-1}$, which corresponds to a $10\%$ relative error. Panel~(b) reports this resulting $\chi$ as a function of $\Ra$. In addition to the cases $\Ra=10^{9}$ and $\Ra=10^{10}$ considered in the previous section, we include $\Ra=10^{8}$ to provide an additional reference point for the dependence on $\Ra$.

Within the explored $\chi$-range, these results suggest a possible scaling advantage of the MPS approach relative to DNS as $\Ra$ increases. The $\chi$ required to recover the global heat transport at fixed accuracy increases only weakly from $\Ra=10^{9}$ to $\Ra=10^{10}$. This contrasts with the \textit{a priori} scalings obtained from DNS snapshot analyses in Section~\ref{sec:scaling_of_complexity_with_Ra_A_priori_B}, where maintaining a fixed $\ell_2$-error indicated that the required $\chi$ would increase substantially (approximately doubling) when moving from $\Ra=10^{9}$ to $\Ra=10^{10}$. By contrast, in the present \textit{a posteriori} analysis based on the dynamical simulations, we find comparable accuracy in the global heat transport at $\chi=120$ for both regimes within the averaging window and error metric used here.
\begin{figure}[!t]
    \centering
    \includegraphics[width=0.98\columnwidth]{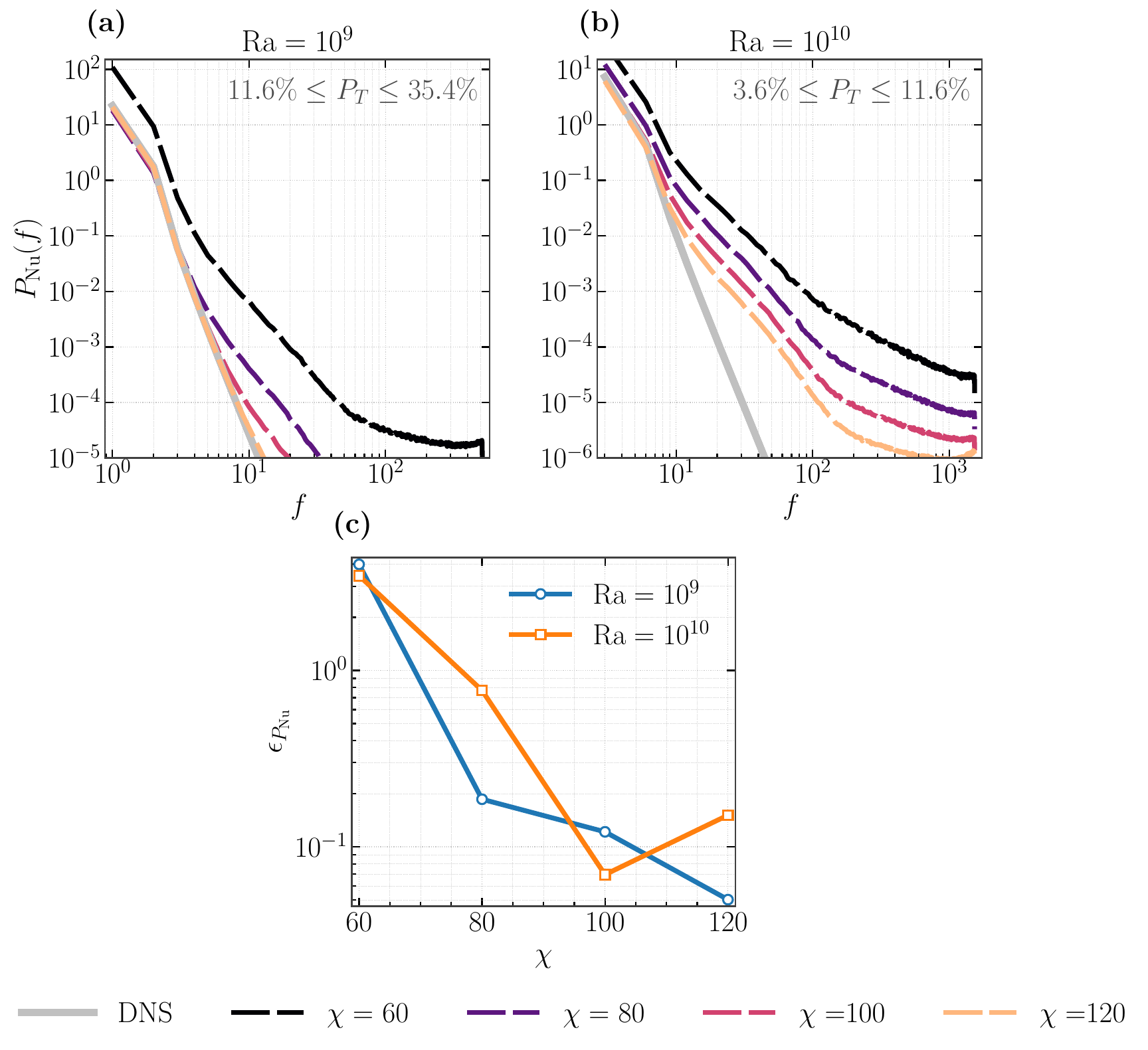}
    \caption{
    \textbf{Temporal spectral analysis of the Nusselt number.} 
    (a,\,b)~Power spectral density $P_{\mathrm{Nu}}(f)$ of the 
    $\mathrm{Nu}_{\mathrm{inst}}(t)$ time series, computed via 
    Welch's method~\cite{Welch1967}, comparing the DNS reference 
    (solid gray) with MPS solutions (dashed) at varying bond 
    dimension~$\chi$ for (a)~$\Ra=10^9$ and (b)~$\Ra=10^{10}$.
    (c)~Relative error in the integrated spectral power 
    $\epsilon_{P_{\mathrm{Nu}}} 
    = |\mathcal{E}_{\mathrm{Nu}}^{\mathrm{MPS}} 
      - \mathcal{E}_{\mathrm{Nu}}^{\mathrm{DNS}}| 
      \,/\, \mathcal{E}_{\mathrm{Nu}}^{\mathrm{DNS}}$ (cf.~Eq.~\ref{eq:energy_error_spectrum})
    as a function of~$\chi$.}
    \label{fig:Nusselt_Welch_spectrum}
\end{figure}
\noindent

\begin{figure*}[t]
    \centering
    \begin{tikzpicture}

        \node[inner sep=0pt] (topPlot) {
                \includegraphics[width=0.96\textwidth, trim={0 0 0cm 0}, clip]{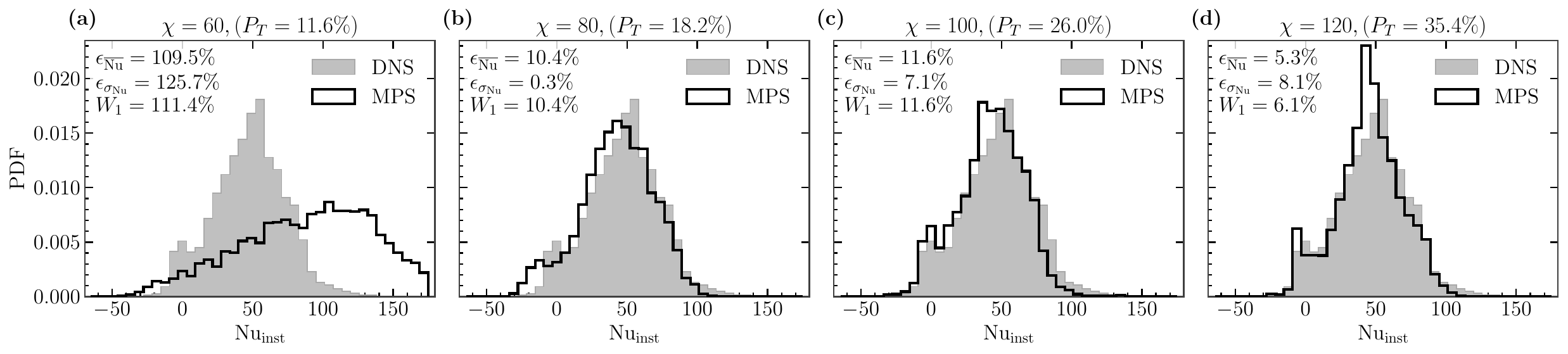}
        };

        \node[inner sep=0pt, below=0.2cm of topPlot] (bottomPlot) {
            \includegraphics[width=0.96\textwidth, trim={0cm 0 0cm 0}, clip]{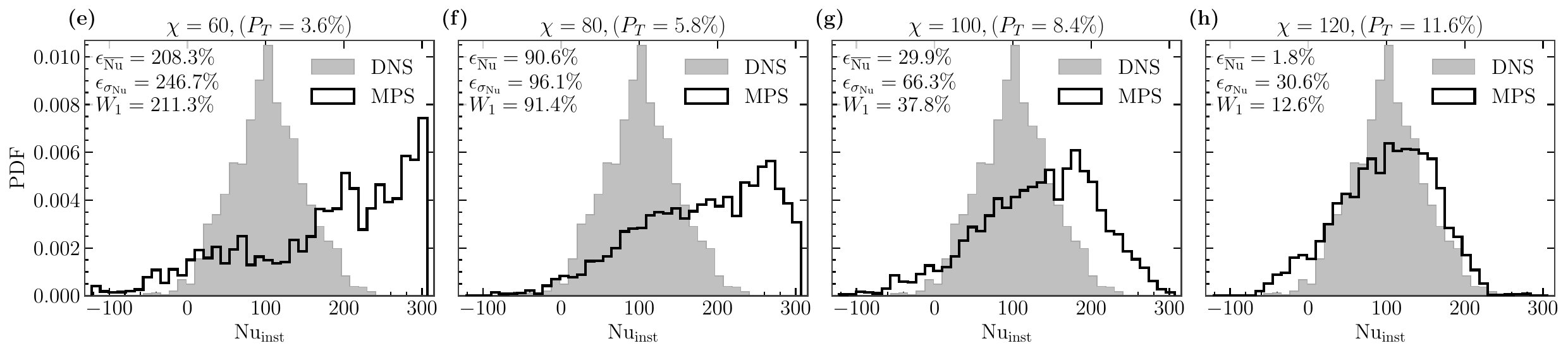}
        };

    \end{tikzpicture}
    \caption{\textbf{Nusselt statistics.} Probability density functions (PDF) of the instantaneous Nusselt number $\text{Nu}_{\text{inst}}$ (Eq.~\eqref{eq:Nu_instantaneous}) for (a–d) $\text{Ra}=10^9$ ($\DT = 24.3\, t_{\ff}$, $M=19$) and (e–h) $\text{Ra}=10^{10}$ ($\DT = 10.0\, t_{\ff}$, $M=24$). %
    Histograms are constructed with bin widths of $6.1$ and $10.6$, respectively, comparing the reference DNS (grey shaded area) with the MPS predictions (solid black lines). Subplots are arranged from left to right by increasing $\chi$, which corresponds to an increasing fraction of parameters $P_T$ (Eq.~\eqref{eq:Pt}) retained by the MPS relative to the DNS. Insets display the relative error in the Nusselt number $\epsilon_{\Nuens}$, the relative error in the standard deviation $\epsilon_{\sigma_{\mathrm{Nu}}}$, and the normalized 1st Wasserstein metric $W_1$ (Eq.~\eqref{eq:W1}).%
    }
    \label{fig:Nusselt_histogram_tikz}
\end{figure*}

Beyond the mean Nusselt number, we examine higher-order 
temporal statistics of the heat transport.
We compute the power spectral density $P_{\mathrm{Nu}}(f)$ of each 
$\mathrm{Nu}_{\mathrm{inst}}(t)$ signal via Welch's 
method~\cite{Welch1967} and average over the ensemble 
(Fig.~\ref{fig:Nusselt_Welch_spectrum}a--b).
Increasing $\chi$ generally improves the agreement with the DNS reference, 
although the convergence is not uniform across all frequencies.
Unlike the sharp cutoffs of Fourier-based filtering~\cite{Boyd2001}, 
which can induce spectral blocking at the truncation limit, the MPS 
spectra transition smoothly from the inertial slope into a bounded 
noise floor.
To summarize the spectral convergence in a single scalar, we define the 
integrated spectral power
\begin{equation}
    \mathcal{E}_{\mathrm{Nu}} 
    = \int_0^{f_{\max}} P_{\mathrm{Nu}}(f) \, \mathrm{d}f \label{eq:energy_error_spectrum}
\end{equation}
and the corresponding relative error 
$\epsilon_{P_{\mathrm{Nu}}} 
= |\mathcal{E}_{\mathrm{Nu}}^{\mathrm{MPS}} 
  - \mathcal{E}_{\mathrm{Nu}}^{\mathrm{DNS}}| 
  \,/\, \mathcal{E}_{\mathrm{Nu}}^{\mathrm{DNS}}$.
Figure~\ref{fig:Nusselt_Welch_spectrum}(c) shows 
$\epsilon_{P_{\mathrm{Nu}}}$ as a function of~$\chi$.
At $\mathrm{Ra}=10^9$, $\epsilon_{P_{\mathrm{Nu}}}$ decreases 
monotonically with~$\chi$.
At $\mathrm{Ra}=10^{10}$, the convergence is non-monotonic: 
$\chi=100$ achieves a lower integrated error than $\chi=120$.
This could be explained by the spectra in panel~(b), where $\chi=120$ 
underpredicts the low-frequency power relative to DNS.
Because the low-frequency modes carry the dominant weight in the integral, 
this underprediction outweighs the improved accuracy at higher frequencies, 
resulting in a larger integrated error.

To complement the spectral analysis, we examine the statistical 
distribution of the instantaneous heat transport.
Figure~\ref{fig:Nusselt_histogram_tikz} displays the PDFs generated 
by pooling the instantaneous samples $\mathrm{Nu}_{\mathrm{inst}}(t)$ 
from all $M$ realizations, quantified by the normalized first 
Wasserstein metric $W_1$ and the relative error in the standard 
deviation $\epsilon_{\sigma_{\mathrm{Nu}}}$.
At $\mathrm{Ra}=10^9$ (Fig.~\ref{fig:Nusselt_histogram_tikz}a--d), 
the PDFs exhibit the same abrupt transition toward the DNS reference 
observed in Fig.~\ref{fig:Nusselt_mean}(a--b).
At the lowest $\chi$ ($\chi=60$), the systematic overestimation 
of heat transport shows as a completely detached distribution shifted 
toward unphysically high values ($W_1 \approx 111.4\%$).
By $\chi=120$, the MPS achieves significant statistical overlap 
with the DNS reference, with the Wasserstein distance reduced to 
$W_1 \approx 6.1\%$ and the mean error to $5.3\%$.
At $\mathrm{Ra}=10^{10}$ (Fig.~\ref{fig:Nusselt_histogram_tikz}e--h), 
$\chi=60$ and $\chi=80$ substantially overpredict the heat transport 
($\epsilon_{\overline{\mathrm{Nu}}} \approx 208.3\%$), 
driving the model toward a more efficient yet unphysical convective state.
An increment to $\chi=100$ initiates a transition toward the DNS reference, 
while $\chi=120$ reduces the mean error to 
$\epsilon_{\overline{\mathrm{Nu}}} \approx 1.8\%$ and aligns the PDF 
with the DNS qualitatively.
The MPS distribution remains broader than the DNS, however, with 
$\epsilon_{\sigma_{\mathrm{Nu}}} \approx 30.6\%$ and 
$W_1 \approx 12.6\%$, indicating that the full variance of the 
fluctuations is not yet resolved at this $\chi$, 
which is to be compared with the spectral analysis above.

\section{Discussion and Outlook}
\label{sec:discussion}
Previous studies exploring MPS applications to computational 
fluid dynamics have focused predominantly on the isothermal, incompressible 
Navier-Stokes equations in spatially homogeneous, doubly-periodic 
configurations such as 2D decaying turbulence, the 3D Taylor-Green vortex, 
and 3D homogeneous isotropic turbulence~\cite{Gourianov2022b, Hoelscher2024, pisoni2025}. 
In these settings, the absence of walls, boundary layers, and thermal 
coupling favors low-rank representations, and \textit{a priori} evaluations 
of DNS snapshots indicated a plateauing of the required bond 
dimension $\chi_{99}$ at high Reynolds numbers. 
Dynamical MPS simulations, however, have thus far been demonstrated for Reynolds numbers 
within the regime where $\chi$ grows as a power law with Re~\cite{Hoelscher2024}.

Rayleigh--B\'{e}nard convection differs from these configurations 
in several fundamental respects.
It is wall-bounded, with thin thermal and velocity boundary layers 
whose thickness decreases with $\Ra$, creating spatial inhomogeneity 
between the near-wall and bulk regions.
The Boussinesq approximation couples the velocity field to an active 
temperature field~$\theta$, adding an additional dynamical field absent 
in isothermal configurations~\cite{VanDerPoel2015,Zhu2018}.
The flow is continuously driven by the imposed temperature gradient, 
sustaining plume dynamics and large-scale circulation throughout the 
simulation.
How these features affect the practicality of the MPS ansatz has, 
to our knowledge, not been previously investigated.

Our \textit{a priori} analysis shows that while the velocity components 
exhibit initial signs of plateauing in the required~$\chi$ with 
increasing~$\Ra$, the temperature field shows no evidence of saturation 
within the investigated range.
Despite this more demanding \textit{a priori} outlook, our dynamical MPS 
simulations demonstrate that mean heat transport and its statistical 
distribution remain robustly captured at moderate~$\chi$, with the 
required $\chi$ growing more slowly with $\Ra$ than the 
\textit{a priori} snapshot analysis would suggest.
Increasing $\chi$ systematically captures more inter-scale
correlations~\cite{Gourianov2022, Gourianov2022b}, as demonstrated 
by the convergence of the mean Nusselt number, its statistical 
distribution, and the progressive recovery of its spectral content.

Reaching higher $\Ra$ within the present framework will require 
addressing the element-wise product, which remains the dominant 
computational bottleneck.
For the values of $\chi$ investigated here, the variational 
implementation proved most robust, but the algorithms introduced in Refs.~\cite{michailidis2025, Meng2026} represent promising alternatives 
with improved asymptotic scaling that offer significant speedups 
at larger~$\chi$.
Adaptive time stepping, operating close to the CFL stability limit, 
offers a complementary route to reducing computational cost.
Incorporating parallelization strategies along the lines of 
Refs.~\cite{Stoudenmire2013, Secular2020} could further accelerate 
time integration, analogous to the domain decomposition employed in conventional DNS~\cite{VanDerPoel2015}. Moreover, TN algorithms map naturally onto GPU architectures~\cite{Hoelscher2024}. In Appendix~\ref{sec:app:Timing_Memory}, we compare the memory scaling of MPS and DNS with increasing $\Ra$, indicating that the compressed format could enable high-$\Ra$ simulations to run entirely within the VRAM of a single GPU.

Future extensions to other architectures, such as tree-tensor networks, provide a route to capturing the anisotropic, multiscale correlations of three-dimensional thermal turbulence. Crucially, this framework also serves as a bridge to quantum computing. By replacing the field ansatz with parameterized variational quantum circuits and compiling differential operators into unitary gate sequences~\cite{Lubasch2020, Jaksch2023, Siegl2026}, our approach establishes a foundation for hybrid quantum-classical simulations of high-$\Ra$ regimes.

\begin{acknowledgments}
This publication and the current work have received funding from the European Union's Horizon Europe research and innovation program (HORIZON-CL4-2021-DIGITAL-EMERGING-02-10) under grant agreement No. 101080085 QCFD. DJ acknowledges the support by DFG project ``Quantencomputing mit neutralen Atomen'' (JA 1793/1-1, Japan-JST-DFG-ASPIRE 2024) and  the Hamburg Quantum Computing Initiative (HQIC) project EFRE. The EFRE project is co-financed by ERDF of the European Union and by ``Fonds of the Hamburg Ministry of Science, Research, Equalities and Districts (BWFGB)''. 
TH and DJ are partly funded by the Cluster of Excellence `Advanced Imaging of Matter' 
of the Deutsche Forschungsgemeinschaft (DFG)|EXC 2056- project ID390715994.
The numerical data is acquired using the computational cluster Hummel-2 of the University of Hamburg, which is funded by the Deutsche Forschungsgemeinschaft (DFG, German Research Foundation) – 498394658. The authors thank P. Siegl for contributions to code development and insightful discussions, and L. Arenstein and G. Marcos for their valuable feedback on the manuscript.
\end{acknowledgments}

\newpage

\bibliography{main.bib}
\clearpage
\newpage

\appendix
\section{Simulation Setup, Initialisation Strategy, and Grid Parameters}
\label{sec:app:apriori_scaling}

All simulations are initialised by first running a base simulation at 
$\Ra = 10^7$ from a linear temperature profile $\theta(y) = -y/L$ with 
zero velocity and pressure fields, superimposing a small sinusoidal 
perturbation of amplitude $\varepsilon = 10^{-7}$ to trigger convective 
instability. Each subsequent simulation at $\Ra^{(m)} \in \{10^8, \ldots, 
10^{11}\}$ is then warm-started from the final state of the converged 
$\Ra^{(m-1)}$ DNS, reducing the spin-up time needed to reach the statistically 
stationary state. The corresponding MPS simulation at each $\Ra$ is 
initialised from the same snapshot, with all fields decomposed into MPS with 
the target $\chi$ via successive SVDs.

Table~\ref{tab:table_1_app} summarises the DNS reference simulations used for
validation in Sections~\ref{sec:MPS_turbulent_fields_A} and~\ref{sec:scaling_of_complexity_with_Ra_A_priori_B}. All simulations use
$\Pr = 1$, aspect ratio $\Gamma=2$, i.e., a cell size of $L \times 2\,L$, and a second-order finite difference scheme on a staggered grid (cf.~Fig.~\ref{fig:staggeredgrid}).

\begin{table}[htbp]
    \centering
    \caption{Summary of Simulation Parameters}
    \label{tab:simulation_summary}
    \resizebox{\columnwidth}{!}{
    \begin{tabular}{|c|c|c|c|c|c|c|c|}
        \hline
        $\Ra$ & $n_x$ & $n_y$ & Time steps & $\delta t$ & $t_{\text{final}}$ & $N_x$ & $N_y$ \\
        \hline
        $10^8$  & 9  & 8  & 60000 & $2.0 \times 10^{-3}$ & 117.19 & 512  & 256  \\
        $10^9$  & 10 & 9  & 30000 & $9.8 \times 10^{-4}$ & 29.30  & 1024 & 512  \\
        $10^{10}$ & 11 & 10 & 75500 & $3.3 \times 10^{-4}$ & 24.58  & 2048 & 1024 \\
        $10^{11}$ & 12 & 11 & 23000 & $1.6 \times 10^{-4}$ & 3.74   & 4096 & 2048 \\
        $10^{12}$ & 13 & 12 & 4000  & $1.2 \times 10^{-4}$ & 0.49   & 8192 & 4096 \\
        \hline
    \end{tabular}}
    \label{tab:table_1_app}
\end{table}

\section{Velocity field compression scaling}
\label{sec:app:v_scaling}

\begin{figure}[htbp]
    \centering
    \includegraphics[width=0.95\columnwidth]{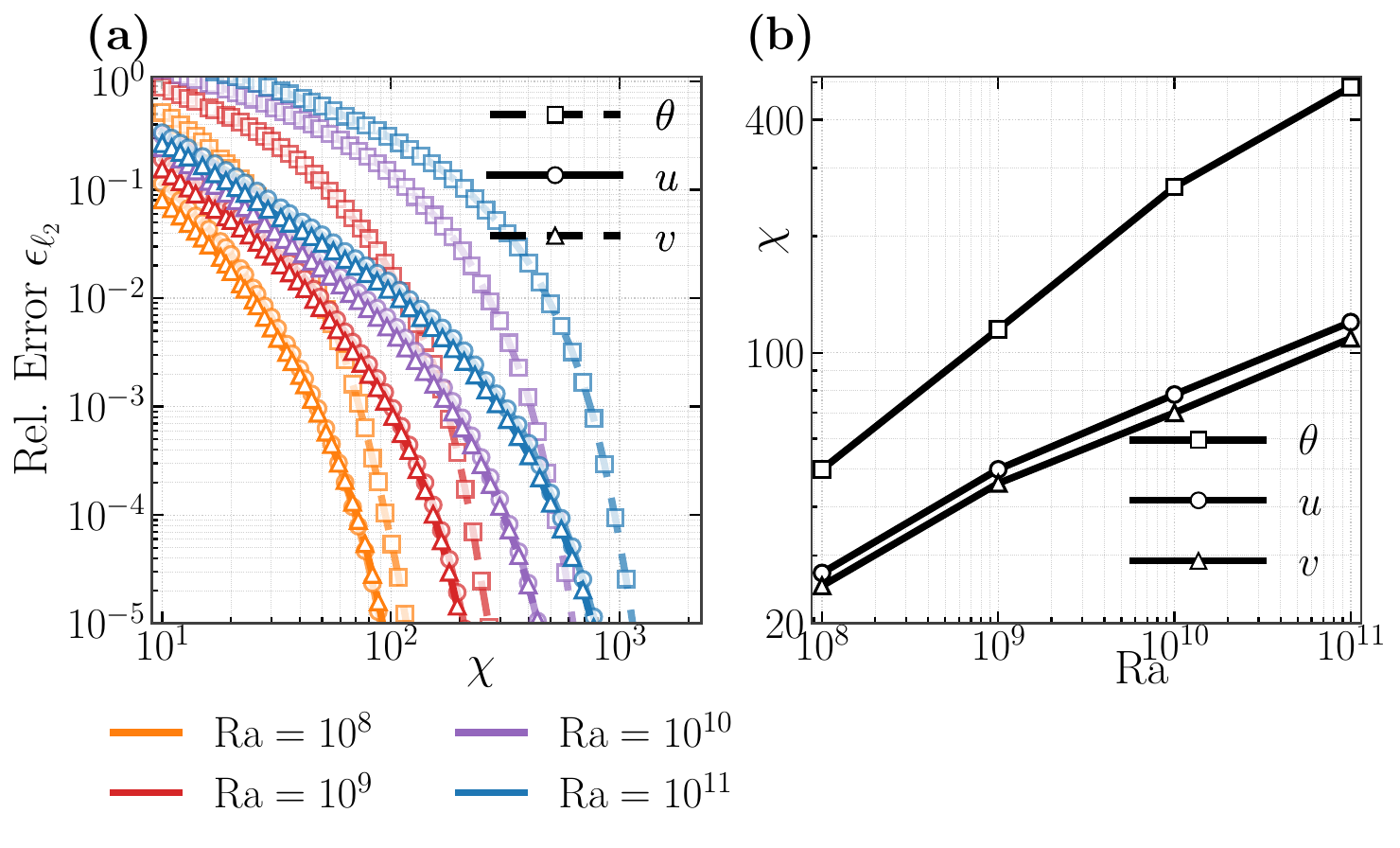}
    \caption{Relative $\ell_2$-error as a function of $\chi$ for $u$, $v$,
    and $\theta$ at various $\Ra$. Compare with Fig.~\ref{fig:scaling}(a);
    the $v$ curves (dashed) nearly overlap with $u$ at all Rayleigh numbers.}
    \label{fig:app:v_scaling}
\end{figure}

Figure~\ref{fig:app:v_scaling} reproduces the a priori compression analysis
of Fig.~\ref{fig:scaling}(a), now including the vertical velocity component $v$.
The $\ell_2$-convergence of $v$ with $\chi$ is virtually indistinguishable from
that of $u$ across all $\Ra$, confirming that both velocity components exhibit
the same attainable compression in the MPS representation.

\section{Data Compression}
\label{sec:app:data_compression}
Table~\ref{tab:nvps_percentage} reports the compression ratio $P_T$ ($N_{\mathrm{MPS}} / N_{\mathrm{full}}$), as given in Eq.~\ref{eq:Pt},
for varying $\Ra$ and $\chi$. Values exceeding 100\% indicate that the MPS representation
requires more parameters than the full state vector, effective compression is achieved only
for $P_T < 100\%$. Note that $P_T$ depends only on the system size $n$ and $\chi$, not on $\Ra$
directly, the $\Ra$ column serves to indicate which system size is used at each Rayleigh number.

\begin{table}[ht]
\centering
\footnotesize
\caption{$P_T$ for Varying $\Ra$ (system sizes) and $\chi$}
\label{tab:nvps_percentage}
\resizebox{\columnwidth}{!}{\begin{tabular}{|l|l|c|c|c|c|c|c|c|}
\hline
 & & \multicolumn{7}{c|}{$\chi$} \\ \cline{3-9}
$\Ra$ & $n$ & 80 & 90 & 100 & 110 & 120 & 150 & 170 \\ \hline
$10^{8}$  & 17 & 53.25\% & 62.99\% & 73.64\% & 85.20\% & 97.69\% & 126.26\% & 143.83\% \\ \hline
$10^{9}$  & 19 & 18.20\% & 21.93\% & 26.04\% & 30.53\% & 35.41\% & 48.73\%  & 58.01\%  \\ \hline
$10^{10}$ & 21 & 5.77\%  & 7.03\%  & 8.42\%  & 9.94\%  & 11.60\% & 16.47\%  & 20.01\%  \\ \hline
$10^{11}$ & 23 & 1.75\%  & 2.14\%  & 2.58\%  & 3.06\%  & 3.59\%  & 5.19\%   & 6.38\%   \\ \hline
\end{tabular}}
\end{table}

\section{Perturbations to Initial Conditions}
\label{app:sec:perturbations}
We first compute the root-mean-square (RMS) fluctuation of the initial temperature field $\theta_0(x, y, t)$ as:
\begin{equation}
    \theta_{\text{rms}} = \sqrt{\langle \theta_0^2 \rangle - \langle \theta_0 \rangle^2},
\end{equation}
where $\langle \cdot \rangle$ denotes the spatial average over the domain area $A = W \times L$.
The spatial structure of the perturbation is constructed to vanish at the walls ($y=\pm L/2$) while introducing low-wavenumber mixing
\begin{equation}
    P(x,y) = \left(\frac{L^2}{4} - y^2\right) \cos\left( \frac{2\pi k x}{W} + \phi \right).
\end{equation}
Here, the wavenumber $k$ is selected uniformly from $\{1, 2, 3\}$ and the phase $\phi$ from $\mathcal{U}[0, 2\pi)$. To ensure zero net heat addition, we remove the spatial mean of the shape function, $P' = P - \langle P \rangle$.
Finally, the perturbation is normalized to a target intensity $\varepsilon$ and superimposed onto the field
\begin{equation}
    \theta_{\text{perturbed}} = \theta_0 + \varepsilon \, \theta_{\text{rms}} \, \frac{P'}{\sigma_{P}},
\end{equation}
where $\sigma_{P} = \sqrt{\langle P'^2 \rangle}$ is the standard deviation of the perturbation shape.

\begin{table}[h!]
    \centering
    \caption{Perturbation intensity $\varepsilon$ applied to the initial temperature field for DNS and MPS for the different $\Ra$.}
    \label{tab:perturbation_params}
    \begin{tabular*}{\columnwidth}{@{\extracolsep{\fill}}lcc@{}}
        \toprule
        $\Ra$ & Method & $\varepsilon$ \\
        \midrule
        $10^8$    & MPS & 5\% \\
                  & DNS & 10\% \\
        \midrule
        $10^9$    & MPS & 1\% \\
                  & DNS & 4\% \\
        \midrule
        $10^{10}$ & MPS & 1\% \\
                  & DNS & 15\% \\
        \bottomrule
    \end{tabular*}
\end{table}

\section{Timing and Memory Requirements}
\label{sec:app:Timing_Memory}
\begin{figure}[!ht]
    \centering
    \includegraphics[width=0.99\columnwidth]{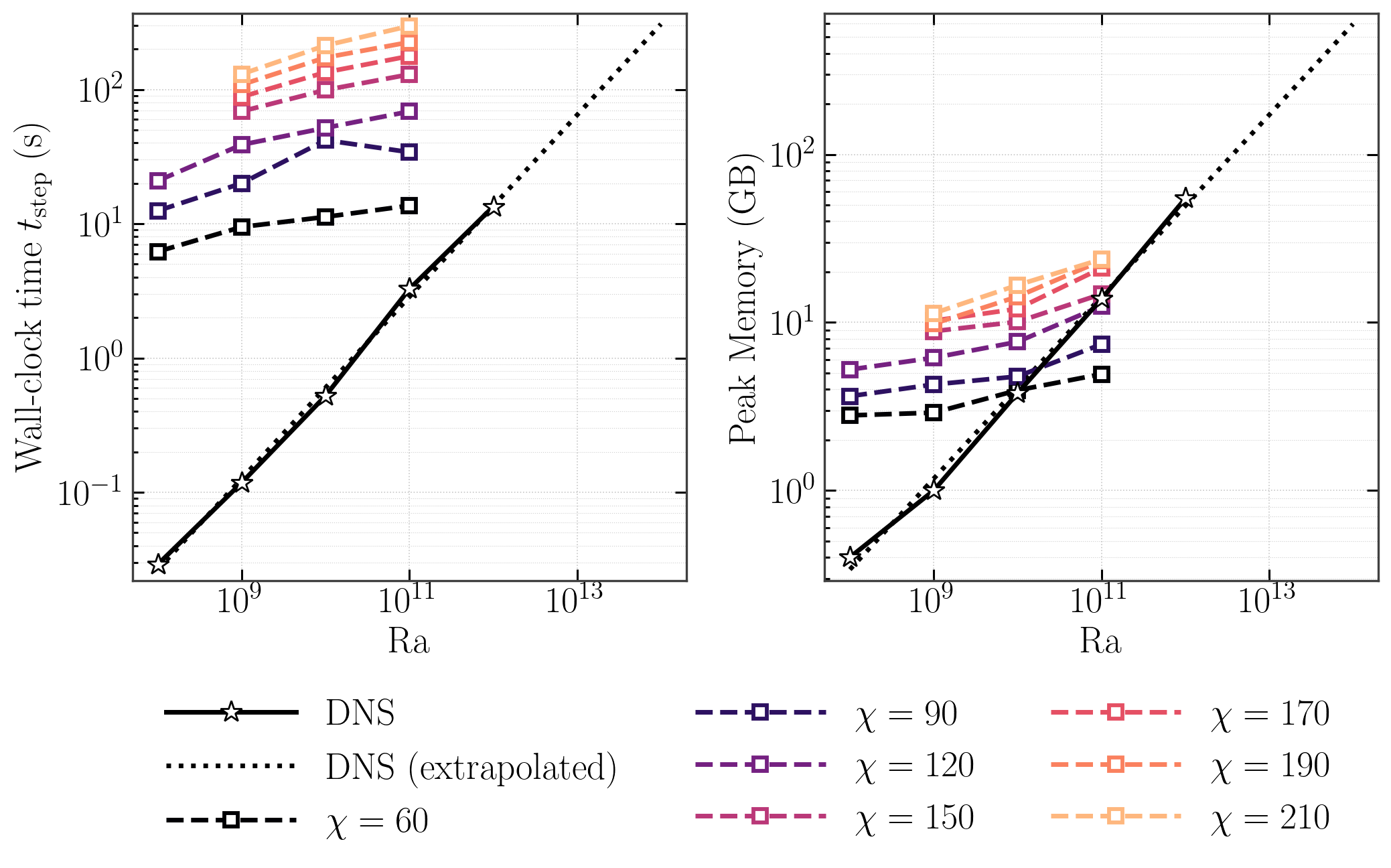}
    \caption{Wall-clock time per timestep $t_{\mathrm{step}}$ (left) and peak memory usage in GB (right) 
as a function of Ra for DNS (solid, stars) and MPS (dashed, squares). MPS results are shown for $\chi \in \{60, 90, \ldots, 210\}$.}
    \label{fig:app:time_and_memory}
\end{figure}

Figure~\ref{fig:app:time_and_memory} reports the wall-clock time per timestep
and peak memory usage for the MPS simulations in Section~\ref{sec:turbulent_statistics_D} at fixed $\chi$ as a function of
Ra, alongside the corresponding DNS measurements.

\section{Statistical Estimation and Error Propagation}
\label{sec:app:Nu_error_prop}
To characterize the convergence of the heat transport statistics, we define the 
time-averaged Nusselt number for each independent realization $m \in \{1, \dots, M\}$ 
over a sampling window $\DT$ as
\begin{equation}
    \mathrm{Nu}_m(\DT) = \frac{1}{\DT} \int_{t_0}^{t_0 + \DT} \mathrm{Nu}_{\text{inst}, m}(t) \, dt \, ,
\end{equation}
where $\mathrm{Nu}_{\text{inst}, m}$ is given by Eq.~\eqref{eq:Nu_instantaneous}. 
The ensemble mean, representing the spatio-temporal average across all $M$ realizations, is then
\begin{equation}
    \Nuens(\DT) = \frac{1}{M} \sum_{m=1}^{M} \mathrm{Nu}_m(\DT) \, .
\end{equation}
The statistical uncertainty of this estimator is quantified by the Standard Error of the Mean (SEM), derived from the variance of the independent temporal histories
\begin{equation}
    \mathrm{SEM}(\DT) = \frac{1}{\sqrt{M}} \sqrt{\frac{1}{M-1} \sum_{m=1}^{M} \left( \mathrm{Nu}_m(\DT) - \Nuens(\DT) \right)^2} \, .
\end{equation}
We define the relative error $\epsilon_{\Nuens}(\DT)$ as
\begin{equation}
    \epsilon_{\Nuens}(\DT) = \frac{\left| \Nuens^\mps(\DT) - \Nuens^\dns(\DT) \right|}{\Nuens^\dns(\DT)} \, ,
\end{equation}
and propagate the uncertainties from both ensembles via the total standard error
\begin{equation}
    \mathrm{SEM}_{\text{tot}} = \sqrt{ \left( \frac{\mathrm{SEM}^\mps }{\Nuens^\dns} \right)^2 + \left( \frac{\Nuens^\mps \cdot \mathrm{SEM}^\dns }{(\Nuens^\dns)^2} \right)^2 } \, ,
\end{equation}
where $\mathrm{SEM}^{\dns}$ and $\mathrm{SEM}^{\mps}$ are the cumulative SEM 
values for the DNS and MPS ensembles, respectively.
\end{document}